\documentclass[twocolumn,superscriptaddress,amsmath, amssymb,prl, aps]{revtex4-1}

\usepackage{graphicx} % Include figure files
\usepackage{dcolumn} % Align table columns on decimal point
\usepackage{bm} % bold math

\usepackage{hyperref} % add hypertext capabilities

\usepackage[linesnumbered, ruled, vlined]{algorithm2e}

\usepackage{xcolor}

\usepackage{amsthm}

\usepackage{xr}
\usepackage{slashed}

\usepackage{tikz}
\usetikzlibrary{quantikz}

\allowdisplaybreaks

\makeatletter
\@addtoreset{equation}{subsection}
\makeatother

\usepackage{MnSymbol}

\usepackage{soul}

\makeatletter
\newcommand{\vast}{\bBigg@{3.0}}
\newcommand{\Vast}{\bBigg@{4.0}}

\makeatother

\usepackage{accents}
\newlength{\dhatheight}

\begin{document}

\preprint{APS/123-QED}

\title{Decoherence Mitigation by Embedding a Logical Qubit in a Qudit}
% \thanks{A footnote to the article title}

\author{Hideyuki Miyahara}
% \altaffiliation[Also at ]{Physics Department, XYZ University.}
\email{miyahara@g.ucla.edu, hmiyahara512@gmail.com}
% \email{miyahara@g.ucla.edu}
% \email{hmiyahara512@gmail.com}

\affiliation{
Department of Electrical and Computer Engineering,
University of California, Los Angeles, CA 90095
}

% \affiliation{
% International Research Center for Neurointelligence (WPI-IRCN), UTIAS,
% The University of Tokyo,
% % 7-3-1 Hongo, Bunkyo-ku,
% Tokyo 113_0033, Japan
% }

% \author{Yasunari Suzuki}
%
% \affiliation{
% NTT Computer and Data Science Laboratories, NTT Corporation, Musashino 180-8585, Japan
% }

\author{Yiyou Chen}
\email{gerry99@ucla.edu}
\affiliation{
Department of Computer Science,
University of California, Los Angeles, CA 90095
}

\author{Vwani Roychowdhury}

\email{vwani@g.ucla.edu}

\affiliation{
Department of Electrical and Computer Engineering,
University of California, Los Angeles, CA 90095
}

\affiliation{
Center for Quantum Science and Engineering, University of California, Los Angeles, CA 90095
}

\author{Louis-S. Bouchard}

\email{louis.bouchard@gmail.com}

\affiliation{
Department of Chemistry and Biochemistry, University of California, Los Angeles, CA 90095
}
\affiliation{
Center for Quantum Science and Engineering, University of California, Los Angeles, CA 90095
}
\affiliation{
California NanoSystems Institute, University of California, Los Angeles, CA 90095
}

\date{\today}

\begin{abstract}
Quantum information stored in a qubit is rapidly lost to the environment. The realization of robust qubits is one of the most important challenges in quantum computing.
Herein we propose to embed a logical qubit within the manifold of a qudit as a scheme to preserve quantum information over extended periods of time. Under identical conditions (e.g. decoherence channels) the submanifold of the logical qubit exhibits extended lifetimes compared to a pure two-level system (qubit).  The retention of quantum information further improves with separation between the sublevels of the logical qubit.  Lifetime enhancement can be understood in terms of entropy production of the encoding and non-encoding subspaces during evolution under a quantum map for a $d$-level system. 
The additional pathways for coherent evolution through intermediate sublevels within a $d$-level manifold provide an information-preserving mechanism: reversible alternative channels to the irreversible loss of information to the environment characteristic of open quantum systems.
\end{abstract}

\maketitle

% \tableofcontents

\section{Introduction}

Schemes for quantum computation (QC) rely on networks of quantum bits (qubits) that are mutually coupled and subjected to sequences of controlled operations (quantum gates).
Each individual qubit is represented by an algebra of zero-trace $2 \times 2$ unitary Hermitian matrices while a classical computer utilizes classical bits that take the value 0 or 1~\cite{Feynman_001, Deutsch_001, Deutsch_002, Lloyd_001}.
The resulting topological manifold describing the state of the qubit is the Bloch sphere, which encodes an uncountably infinite number of possible states, as compared to the two discrete states of a classical bit.
Multi-qubit states are described in terms of tensor products of Bloch spheres.
The richness of information encoded by qubits endow quantum computers with unique computational capabilities.
Early on, Shor~\cite{Shor_001, Fowler_002} Grover~\cite{Grover_001} proposed algorithms that have almost become synonymous with QC.
Unfortunately, the creation of clean quantum gates exhibiting high fidelity in the presence of decoherence have proven to be a formidable challenge.
Quantum error correction (QEC) codes were proposed to mitigate these problems~\cite{Shor_002, Nielsen_001}.
The highly important toric and surface codes have had a great impact in the field~\cite{Bravyi_001, Kitaev_001, Kitaev_002, Bravyi_002, Nielsen_001, Steane_001, Devitt_001, Fowler_001}.
Improvements in quantum technology over the last two decades have led to the successful realization of the surface code~\cite{Johnson_001, Arute_001}; see also~\cite{Krinner_001, Zhao_001}.

There currently exist several different schemes for the realization of physical qubits (e.g. molecular magnets~\cite{Leuenberger_001}, defects in solids, cold ions, atoms and molecules, Rydberg atoms, nuclear and electron spin resonances to name a few), superconducting qubit platforms are currently the most popular thanks to their public availability on cloud computing platforms made possible by companies such as Google, IBM, Rigetti, IonQ and so on~\cite{Castelvecchi_001}.
These platforms are termed NISQ (noisy intermediate scale quantum) devices.
As pointed out by Preskill~\cite{Preskill_001}, NISQ devices have found many immediate applications, such as  variational quantum algorithms~\cite{Cerezo_001, Mitarai_001, Schuld_001, Miyahara_001} and quantum-inspired algorithms~\cite{Han_001,Arrazola_001, Miyahara_002, Miyahara_003}.
As of today, it is unclear whether any of these methods provide a clear quantum advantage over classical computers in applications of current interest.
We note, on the other hand, that the feasibility of quantum supremacy has been demonstrated in experiments~\cite{Arute_001}.

Indeed, while these NISQ platforms provide excellent research, development and educational tools, the realization of large-scale QC algorithms remains out of reach, as the qubits remain too fragile, not well-controlled, and limited by decoherence as well as leakage effects.
The impact of quantum errors can be mitigated using error-correction schemes.
However, those are based on fault-tolerant architectures, which require additional hardware resources (qubits and related controls).
The added overhead limits the usefulness of error correction schemes.
To overcome this problem, physical devices that are inherently fault tolerant have been proposed as a way to overcome logical errors~\cite{Szkopek_001}.
With physical fault tolerance, no special  architectures are needed, as the fault tolerance is built into the design of the qubit itself.
This circumvents the problem of different setups producing different sources of errors.
In this context, the realization of robust qubits is of great interest and utmost importance.

In the context of generating robust qubits we note the proposals for multi-level systems, such as error-correcting codes for molecular magnets~\cite{Chiesa_001, Chiesa_002}, optical systems~\cite{Puri_001} as well as the Kerr-cat qubit (superposition of coherent states)~\cite{Grimm_001}.
In Ref.~\cite{Tacchino_001} the authors proposed molecular spin qudits as elements of a quantum simulator.
In Ref.~\cite{Chalopin_001}  dysprosium (Dy), whose spin $J$ is 8, was used for enhanced quantum sensing.
Furthermore, bosonic codes have been extensively studied: the binomial code~\cite{Michael_001}, the cat code~\cite{Leghtas_001, Bergmann_001, Mirrahimi_001, Li_001, Hastrup_001}, the Gottesman-Kitaev-Preskill (GKP) code~\cite{Gottesman_001, Grimm_001, Tzitrin_001, Bourassa_001, Larsen_001}, and the rotation-symmetric bosonic codes~\cite{Grimsmo_002}.
To realize the bosonic codes, truncated bosonic modes are required and optical systems are possible platforms that may provide us with  bosonic modes.
Nuclear spins with  $I > 1/2$ are examples of multi-level spin systems, they exhibit faster relaxation rates, which are a function of the nuclear quadrupole moment and local electric-field gradient.
In Ref.~\cite{Sanctuary_001, Zur_001, Vega_001, Vega_002}, the structure and control of such systems is analyzed in the context of NMR experiments.

The description of qudit systems and their uses in QC is often done in terms of generalized Pauli and Clifford operators~\cite{Hostens_001}, qudit surface code~\cite{Bullock_2007}, a decoder for the qudit surface code~\cite{Watson_001}, and a quantum error correction architecture for qudits~\cite{Nadkarni_001, Klappenecker_001}.
Finally, in Ref.~\cite{Wang_001} qudit-based QC is discussed, including implementations of qudit variants of known quantum algorithms.
In this study, we examine the case of qudits ($d$-level systems) that encode quantum states in a superposition of two energy sublevels playing the role(s) of logical qubits.
The remaining unused levels (the ``non-encoding subspace'') are not explicitly addressed.
Instead, this subspace naturally enhances the lifetime of the logical qubits. We focus on the spin-lattice relaxation mechanism, where enhanced relaxation motivates the use of such systems in quantum memory applications.
Specifically, we show that when the quantum information is stored into the two most polarized spin states (e.g., $\ket{\psi} = c_1 \ket{d-1} + c_2 \ket{0}$) one obtains a more robust quantum memory in which quantum information can be stored for longer periods of time compared to a pure two-level system (qubit) with no intermediate sublevels separating the pair of qubit levels.
In numerical computations we show that the lifetime of the proposed quantum memory is longer than the conventional qubit system.
In order to understand the flow of information between encoding and non-encoding subspaces, entropy production provides an explanation for why qudits exhibit fundamentally different behavior compared to qubits.  
% We explore the possibility of qudits as quantum memory and/or a quantum code.
This paper is organized as follows.
In Secs.~\ref{main_sec_qudit_001_001}, \ref{main_sec_process_fidelity_001_001}, and \ref{main_sec_entropy_production_001_001}, we describe the mathematical prerequisite.
In Secs.~\ref{main_sec_numerical_simulations_fidelity_001_001} and \ref{main_sec_numerical_simulations_entropy_production_001_001}, we show numerical simulations and give discussions.
Finally, Sec.~\ref{main_conclusions_001_001} concludes this paper.   In Appendix we compare the qudit embedding method to a simple QEC model and identify a potentially interesting direction for future research, namely, the use of qudits in quantum memory applications with inherent robustness to information loss.

\section{Maximally polarized states (spin-coherent states) in qudits} \label{main_sec_qudit_001_001}

In this section we describe how to embed qubit information into a qudit system.
Consider a qudit with spin $s$ whose Hilbert space is spanned by $d = 2s + 1$ energy levels.
The density matrix for a qubit $\hat{\rho}'$ is specified via two basis states $| \uparrow \rangle, | \downarrow \rangle$ and four complex numbers $\rho_{\uparrow \uparrow}, \rho_{\uparrow \downarrow}, \rho_{\downarrow \uparrow}, \rho_{\downarrow \downarrow}$ that satisfy $\rho_{\uparrow \uparrow} + \rho_{\downarrow \downarrow} = 1$
%$\rho_{\uparrow \uparrow}^* = \rho_{\uparrow \uparrow}$, $\rho_{\downarrow \downarrow}^* = \rho_{\downarrow \downarrow}$,
and $\rho_{\uparrow \downarrow}^* = \rho_{\downarrow \uparrow}$:
\begin{align}
\hat{\rho}' &\coloneqq \rho_{\uparrow \uparrow} | \uparrow \rangle \langle \uparrow | + \rho_{\uparrow \downarrow} | \uparrow \rangle \langle \downarrow | + \rho_{\downarrow \uparrow} | \downarrow \rangle \langle \uparrow | + \rho_{\downarrow \downarrow} | \downarrow \rangle \langle \downarrow |. \label{main_qubit_state_001_001}
\end{align}
We denote a single-{\it qudit} state by $\hat{\rho}$ and the $d$ states in the $d$-level qudit by $| 0 \rangle, | 1 \rangle, | 2 \rangle, \dots, | d - 1 \rangle$. The state described in Eq.~\eqref{main_qubit_state_001_001} is encoded into the $d$-level qudit as follows:
\begin{align}
\hat{\rho} &\coloneqq \rho_{\uparrow \uparrow} | 0 \rangle \langle 0 | + \rho_{\uparrow \downarrow} | 0 \rangle \langle d-1 | \nonumber \\
& \quad + \rho_{\downarrow \uparrow} | d-1 \rangle \langle 0 | + \rho_{\downarrow \downarrow} | d-1 \rangle \langle d-1 |. \label{main_qudit_state_001_001}
\end{align}
Note that $| 0 \rangle$ and $ | d-1 \rangle$ are often referred to as $| s, s \rangle$ and $| s, -s \rangle$, respectively.  They are also called ``maximally polarized'' states or spin-coherent states.
In this paper, we investigate some properties of Eq.~\eqref{main_qudit_state_001_001}, such as its lifetime by using an error model that will be explained below.  Our analysis shows that encoding of the qubit state in the maximall polarized states leads to  longer lifetimes compared to the use of  intermediate levels.

\section{Error models} \label{main_sec_error_model_001_001}

We start by defining the generalized Pauli operators~\cite{Gheorghiu_001}:
\begin{align}
 \hat{X}' &\coloneqq \sum_{k \in \mathbb{F}_{d-1}} (| k \rangle \langle k + 1 | + | k + 1 \rangle \langle k |). \label{main_generalized_Pauli_Xp_001_001} \\
  \hat{Z} &\coloneqq \sum_{k \in \mathbb{F}_d} \omega^k | k \rangle \langle k |, \label{main_generalized_Pauli_Z_001_001}
\end{align}
where $\mathbb{F}_d \coloneqq \{0, 1, 2, \dots, d-1\}$, $\omega \coloneqq e^{2 \pi i / d}$, and $| d \rangle \coloneqq | 0 \rangle$.
Note that the definition of $\hat{X}'$ in Eq.~\eqref{main_generalized_Pauli_Xp_001_001} is different from the definition of $\hat{X}$ in Ref.~\cite{Gheorghiu_001} since $\hat{X}$ is not symmetric.
For a $n$-qudit system, $\hat{X}'$ in Eq.~\eqref{main_generalized_Pauli_Xp_001_001} and $\hat{Z}$ in Eq.~\eqref{main_generalized_Pauli_Z_001_001} acting on the $i$-th qudit are denoted by
\begin{align}
  \hat{X}_i' &\coloneqq \underbrace{\hat{1} \otimes \dots \otimes \hat{1}}_{i-1} \otimes \hat{X}' \otimes \underbrace{\hat{1} \otimes \dots \otimes \hat{1}}_{n-i}, \\
  \hat{Z}_i &\coloneqq \underbrace{\hat{1} \otimes \dots \otimes \hat{1}}_{i-1} \otimes \hat{Z} \otimes \underbrace{\hat{1} \otimes \dots \otimes \hat{1}}_{n-i}.
\end{align}
Evolution of open quantum systems is continuous in time, as described by a master equation such as the Lindblad equation~\cite{Breuer_001}.  However, in the QEC literature it is customary to consider a discrete-time dynamical model called the quantum map:
\begin{align}
  \hat{\rho}_{t + \Delta t} &= \mathcal{E}_p (\hat{\rho}_t), \label{main_quantum_map_001_001}
\end{align}
where $\Delta t$ is a discrete time step taken to be $\Delta t=1$ (without loss of generality).
For $\mathcal{E}_p (\cdot)$ in Eq.~\eqref{main_quantum_map_001_001}, we consider the following model:
\begin{align}
  \mathcal{E}_p (\cdot) &\coloneqq \bigotimes_{i = 1, 2, \dots, n} \mathcal{E}_p (\cdot; \{ \hat{U}_i^k \}_{k=1}^{d-1}), \label{main_error_model_001_001}
\end{align}
where
\begin{align}
& \bigotimes_{i = 1, 2, \dots, n} \mathcal{E}_p (\cdot; \{ \hat{U}_{i, k} \}_{k=1}^K) \nonumber \\
& \quad \coloneqq \mathcal{E}_p (\mathcal{E}_p ( \dots (\mathcal{E}_p (\cdot; \{ \hat{U}_{n, k} \}_{k=1}^K) \dots; \{ \hat{U}_{2, k} \}_{k=1}^K); \{ \hat{U}_{1, k} \}_{k=1}^K),
\end{align}
and
\begin{align}
\mathcal{E}_p (\hat{\rho}; \{ \hat{U}_{i, k} \}_{k=1}^K) &\coloneqq (1 - p) \hat{\rho} + \frac{p}{K} \sum_{k=1}^K \frac{\hat{U}_{i, k} \hat{\rho} \hat{U}_{i, k}^\dagger}{\mathrm{Tr} [\hat{U}_{i, k} \hat{\rho} \hat{U}_{i, k}^\dagger]}.
\label{main_quantum_map_001_001}
\end{align}
Here, $p$ is the  probability of error in the model and $K$ is the number of error channels.  For the $X'$-type error model and the $Z$-type error model, $K=1$, but for the $X'+Z$-type error model, $K$ = 2.
Single-qudit unitary operators, $\{ \hat{U}_k \}_{k=1}^K$, are denoted as:
\begin{align}
\hat{U}_{i, k} &\coloneqq \underbrace{\hat{1} \otimes \dots \otimes \hat{1}}_{i-1} \otimes \hat{U}_k \otimes \underbrace{\hat{1} \otimes \dots \otimes \hat{1}}_{n-i}.
\end{align}
Note that $\bigotimes_{i = 1, 2, \dots, n} \mathcal{E}_p (\cdot; \{ \hat{U}_{i, k} \}_{k=1}^K)$ does not depend on the order of terms in the product since we consider only single-qudit errors.

\section{Process fidelity} \label{main_sec_process_fidelity_001_001}

A variety of measures~\cite{Liang_001} exist to quantify the time-evolution of quantum states.
We use the process fidelity~\cite{Gilchrist_001, Mayer_001} since our calculations do not invoke a random number generator (e.g., the Haar measure) and its interpretation is relatively clear.
The definition of process fidelity is based on the concept of quantum state fidelity for the maximally entangled state.
The fidelity between two quantum states $\hat{\rho}$ and $\hat{\sigma}$ is given by~\cite{Liang_001}
\begin{align}
  \mathcal{F} (\hat{\rho}, \hat{\sigma}) &\coloneqq \bigg( \mathrm{Tr} \bigg[ \sqrt{\sqrt{\hat{\rho}} \hat{\sigma} \sqrt{\hat{\rho}}} \bigg] \bigg)^2. \label{main_fidelity_001_001}
\end{align}
The initial state is
\begin{align}
  | \psi_\mathrm{ini} \rangle &= \frac{1}{\sqrt{N}} \sum_{i = 1}^N | i \rangle \otimes | i \rangle,
\end{align}
whereas error models are applied to the second qudit.
We compute the fidelity between the initial state and the state at time step $\tau$.

The general definition of the process fidelity is given by~\cite{Gilchrist_001, Mayer_001}
\begin{align}
  \mathcal{F} (\mathcal{A} (\cdot), \mathcal{B} (\cdot)) &\coloneqq \mathcal{F} (\hat{\rho}_\mathcal{A}, \hat{\rho}_\mathcal{B}'), \label{main_def_process_fidelity_001_001}
\end{align}
where $\hat{\rho}_\mathcal{A} \coloneqq \mathcal{A} (\hat{\rho})$ and $\hat{\rho}_\mathcal{B}' \coloneqq \mathcal{B} (\hat{\rho}')$.
In this study, we use the identity operator for $\mathcal{A}$ in Eq.~\eqref{main_def_process_fidelity_001_001} and the error models defined in Sec.~\ref{main_sec_error_model_001_001}.

\section{Entropy production} \label{main_sec_entropy_production_001_001}

To interpret the results from numerical simulations, we used entropy production, which his defined as
\begin{align}
  \Delta S_\tau &\coloneqq S_\tau - S_{\tau - 1}, \label{main_def_entropy_production_001_001}
\end{align}
with $S_{-1} = 0$ and
$S \coloneqq - \mathrm{Tr} [\hat{\rho} \ln \hat{\rho}]$. Furthermore, we define entropy productions in the total space, the encoding subspace, and the non-encoding subspace in terms Eq.~\eqref{main_def_entropy_production_001_001} with $\hat{\rho}$, $\hat{\rho}^\mathrm{en}$, and $\hat{\rho}^\mathrm{non-en}$, respectively, where
\begin{align}
\hat{\rho}^\mathrm{en} &\coloneqq \hat{P}^\mathrm{en} \hat{\rho} \hat{P}^\mathrm{en}, \label{main_rho_encoding_001_001} \\
\hat{\rho}^\mathrm{non-en} &\coloneqq \hat{Q}^\mathrm{en} \hat{\rho} \hat{Q}^\mathrm{en}. \label{main_rho_non-encoding_001_001}
\end{align}
Here $\hat{P}^\mathrm{en}$ is the projection onto the most polarized states $| 0_\mathrm{L} \rangle$ and $| 1_\mathrm{L} \rangle$ whereas $\hat{Q}^\mathrm{en}$ is the projection in the orthogonal complement of $\hat{P}^\mathrm{en}$ (i.e. the non-encoding subspace):
\begin{align}
\hat{P}^\mathrm{en} &\coloneqq (| 0_\mathrm{L} \rangle \langle 0_\mathrm{L} | +| 1_\mathrm{L} \rangle \langle 1_\mathrm{L} |) \otimes (| 0_\mathrm{L} \rangle \langle 0_\mathrm{L} | +| 1_\mathrm{L} \rangle \langle 1_\mathrm{L} |), \\
\hat{Q}^\mathrm{en} &\coloneqq \hat{1} - \hat{P}^\mathrm{en}.
\end{align}

\section{Numerical simulation: fidelity} \label{main_sec_numerical_simulations_fidelity_001_001}

\begin{figure}[t]
\centering
\includegraphics[scale=0.400]{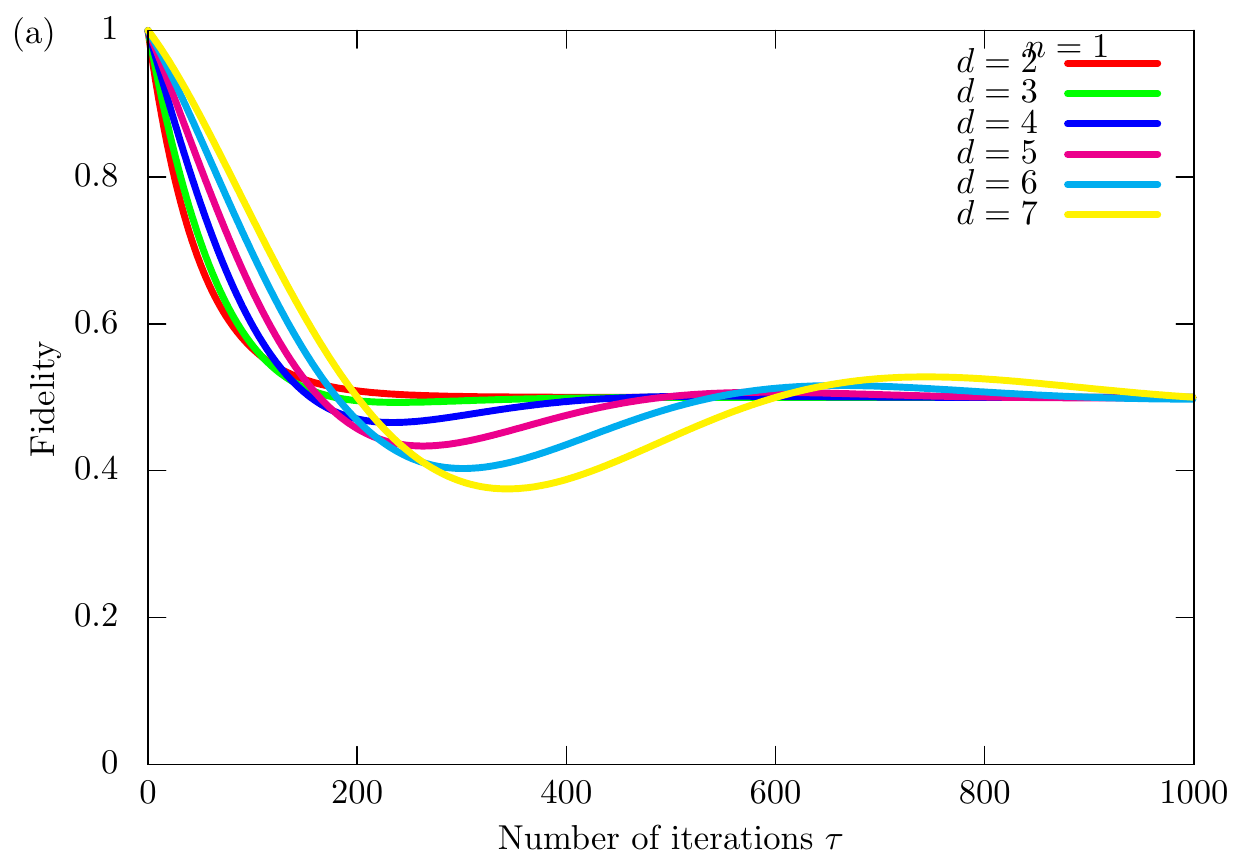}
\includegraphics[scale=0.400]{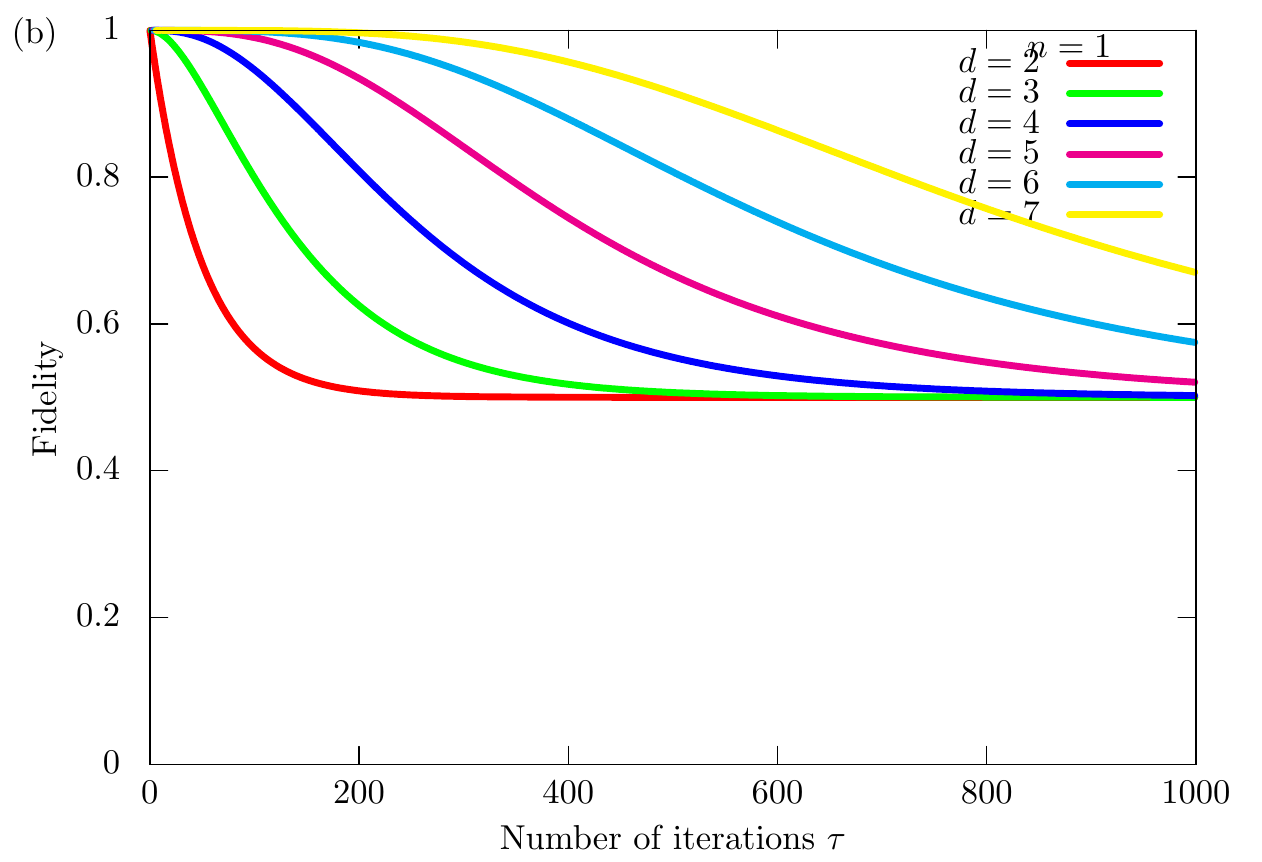}
\includegraphics[scale=0.400]{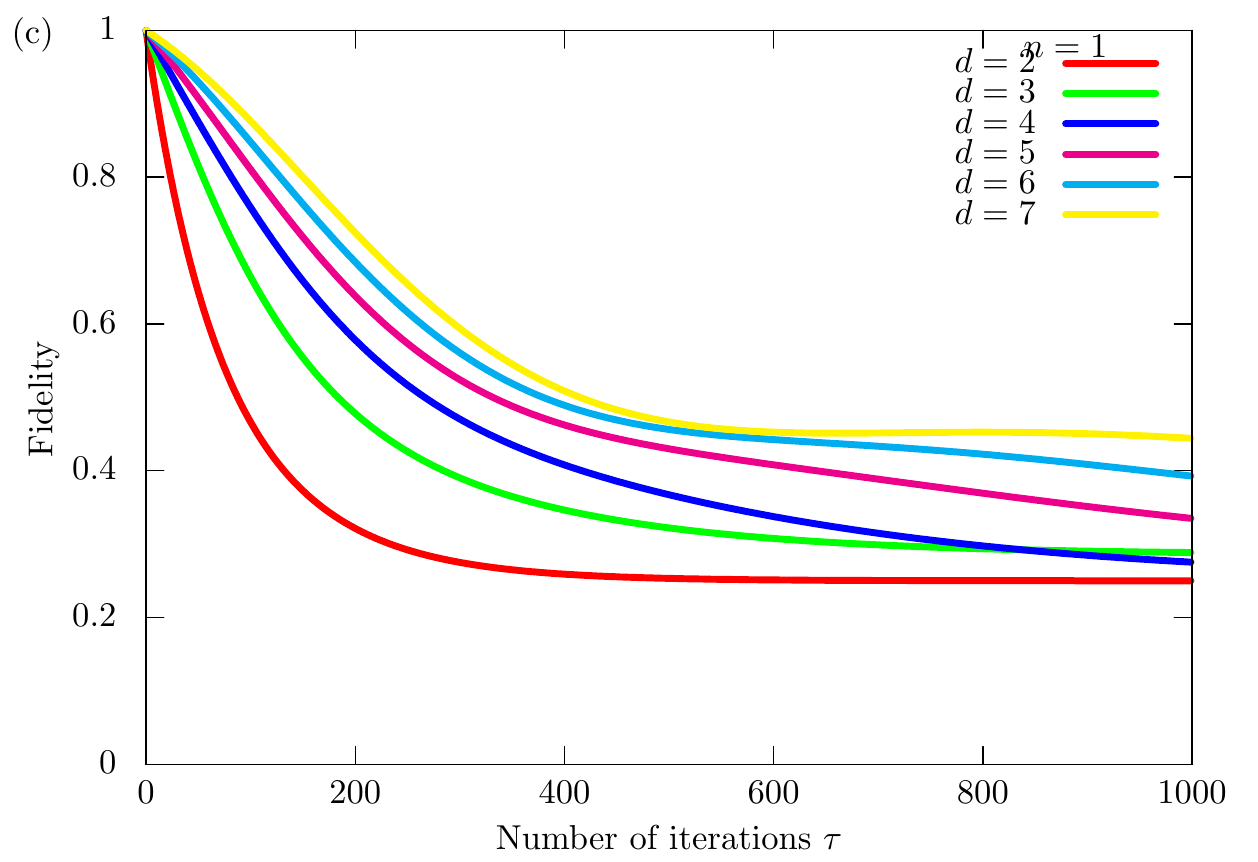}
\caption{Process fidelity for (a) the $Z$-type error model, (b) the $X'$-type error model, and (c) the $X'+Z$-type error model. $d$ is varied. The initial state is the maximally entangled state. The process fidelity decays slower for higher $d$ and this observation means that qudits with higher $d$ show longer life times.}
\label{main_fidelity_000_001}
\end{figure}

To quantify decoherence of quantum states we use the process fidelity, Eq.~\eqref{main_fidelity_001_001} by computing the quantity:
\begin{align}
  \mathcal{F} \bigg(\hat{\rho}_\mathrm{ini}, \frac{\hat{\rho}_t^\mathrm{en}}{\mathrm{Tr} [\hat{\rho}_t^\mathrm{en}]} \bigg),
\end{align}
where $\hat{\rho}_t^\mathrm{en} \coloneqq \hat{P}^\mathrm{en} \hat{\rho}_t \hat{P}^\mathrm{en}$ and $\hat{\rho}_t$ is defined, recursively, via $\hat{\rho}_t = \mathcal{E}_p (\hat{\rho}_{t-1})$.
We denote the actions of the qudit operators $\hat{X}_i', \hat{Z}_i$ and $\hat{X}_i'+ \hat{Z}_i$
in Eq.~\eqref{main_error_model_001_001} by $\mathcal{E}_p^{(X')} (\cdot)$, $\mathcal{E}_p^{(Z)} (\cdot)$, and $\mathcal{E}_p^{(X'+Z)} (\cdot)$, respectively.
Figure~\ref{main_fidelity_000_001} shows process fidelity for the $Z$-,  $X'$-, and $X'+Z$-type error models and its dependence on $d$. Here we take $| 0_\mathrm{L} \rangle = | 0 \rangle$ and $| 1_\mathrm{L} \rangle = | d-1 \rangle$ (maximally polarized state) and increase the dimensionality of the qudit manifold ($d$).
For all three error models process fidelity increases with $d$, at least initially.
Thus, quantum information is preserved over longer times. The logical qubit's lifetime appears to increase with its dimensionality ($d$).  It is still unclear whether this is due to dimensionality or separation between the levels of the logical qudit.

%We have seen that the process fidelity decays slower for larger $d$ when the most polarized states are used.

\begin{figure}[t]
\centering
\includegraphics[scale=0.400]{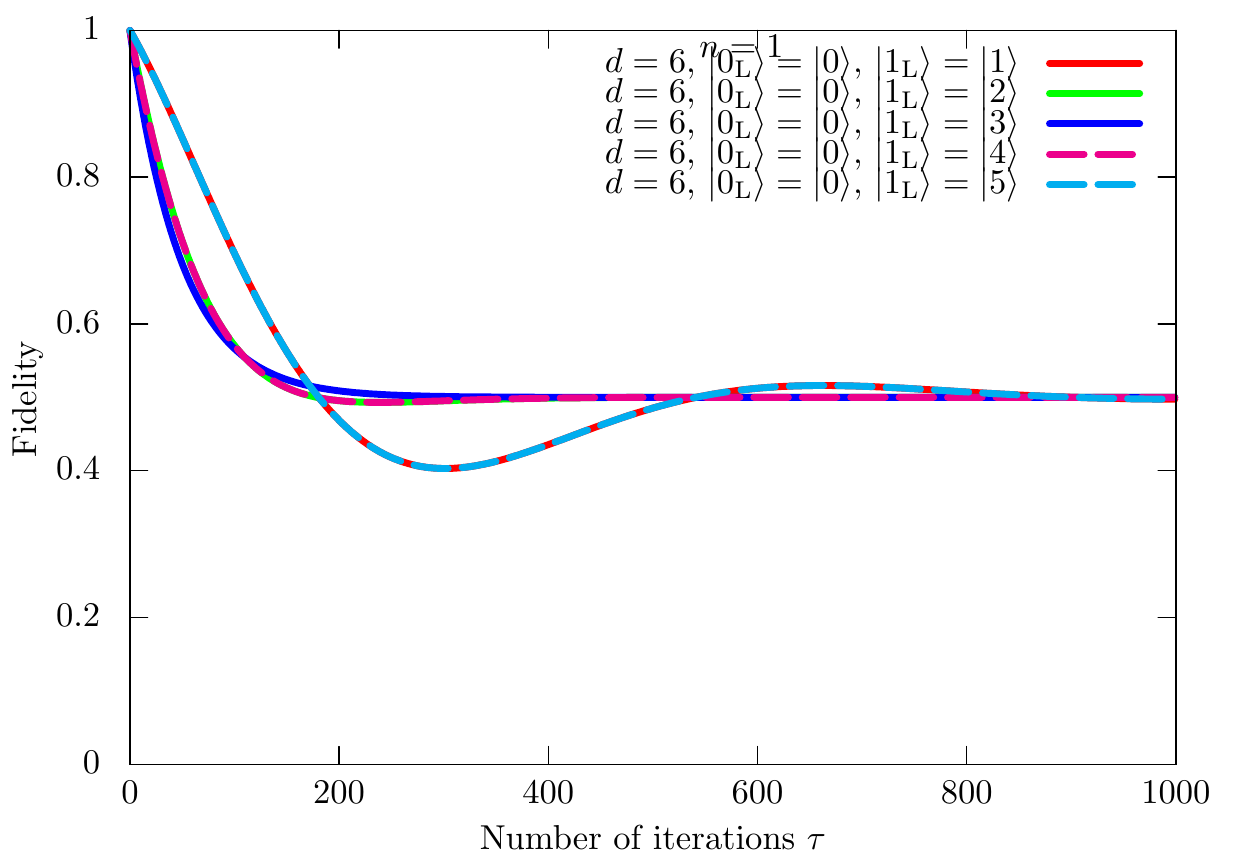}
\includegraphics[scale=0.400]{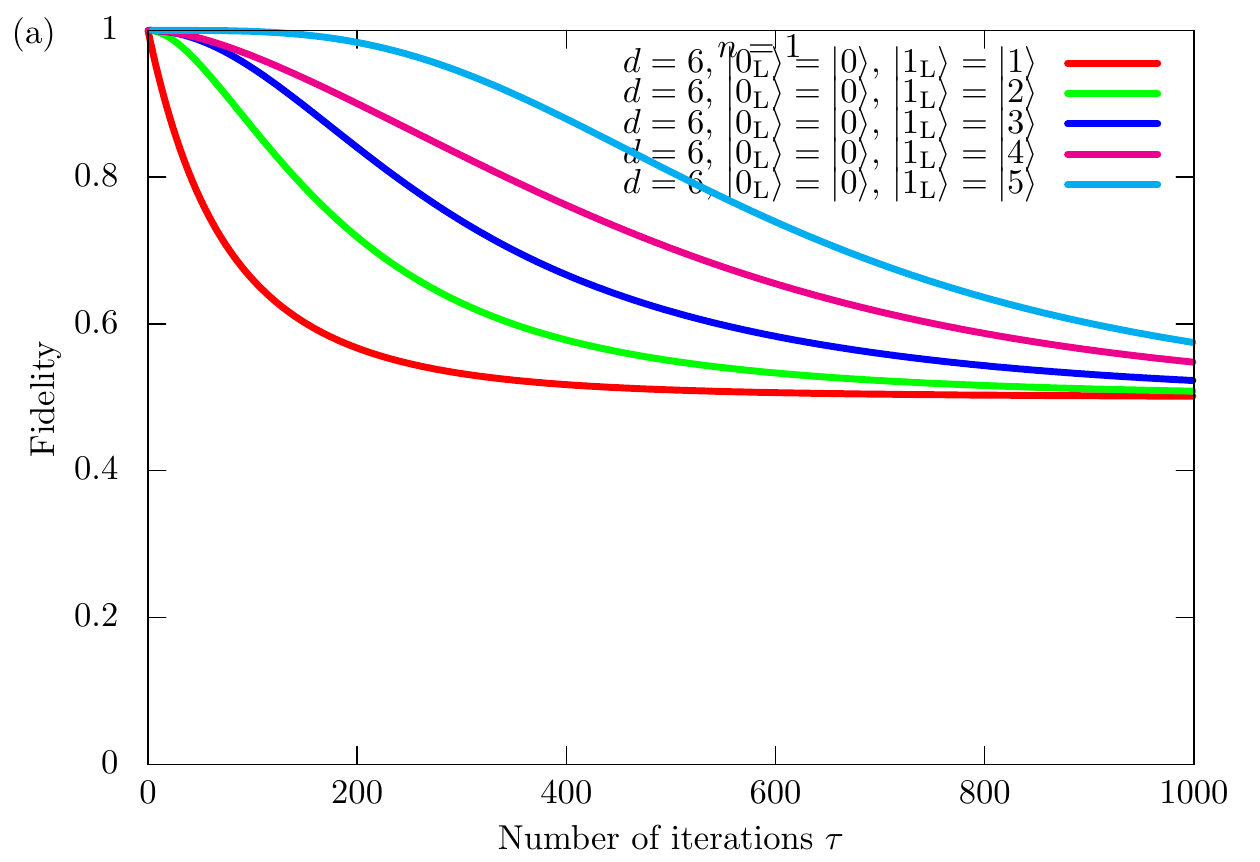}
\includegraphics[scale=0.400]{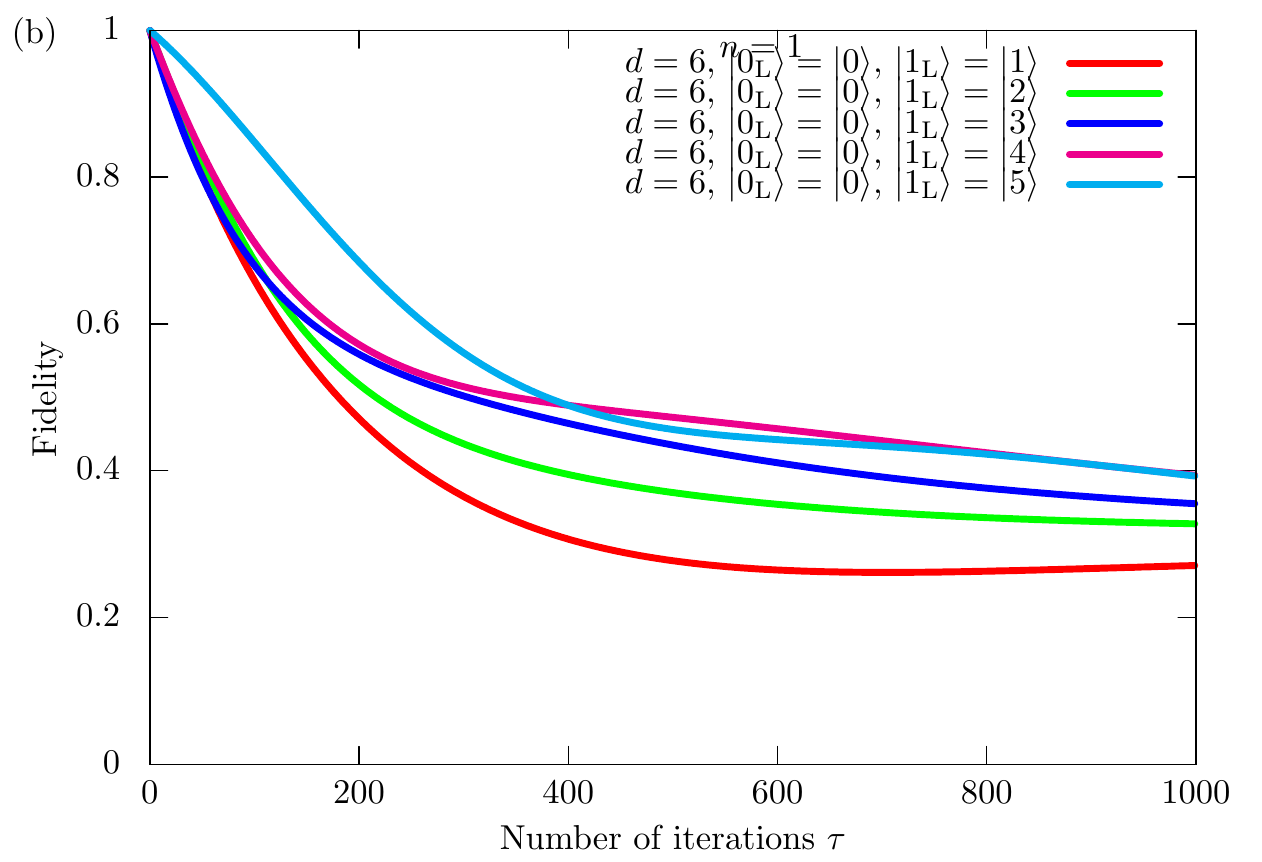}
\caption{Process fidelity for (a) the $Z$-type error model, (b) the $X'$-type error model, and (c) the $X'+Z$-type error model. We have varied $| 1_\mathrm{L} \rangle$ from $| 1 \rangle$ to $| 5 \rangle$. The initial state is the maximally entangled state.}
\label{main_fidelity_000_002}
\end{figure}

\begin{figure}[t]
\centering
\includegraphics[scale=0.400]{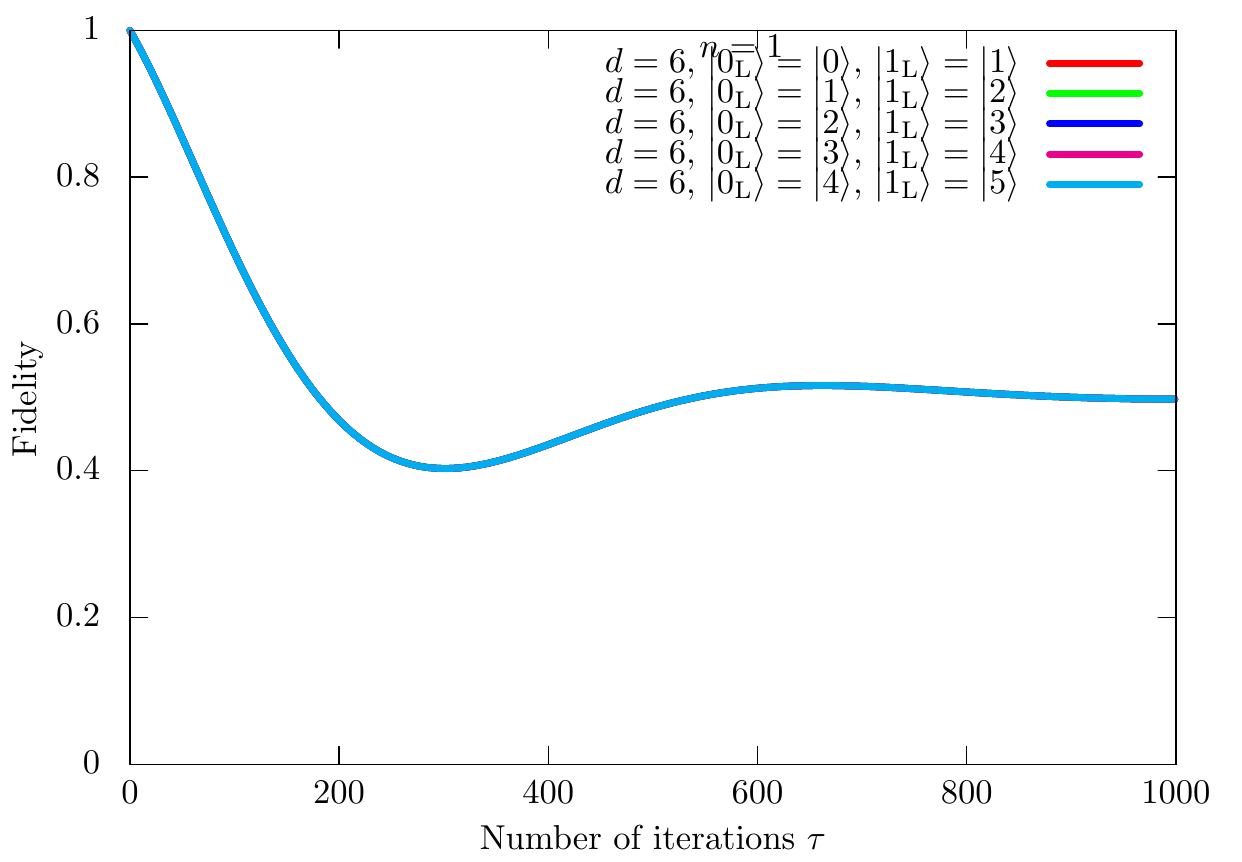}
\includegraphics[scale=0.400]{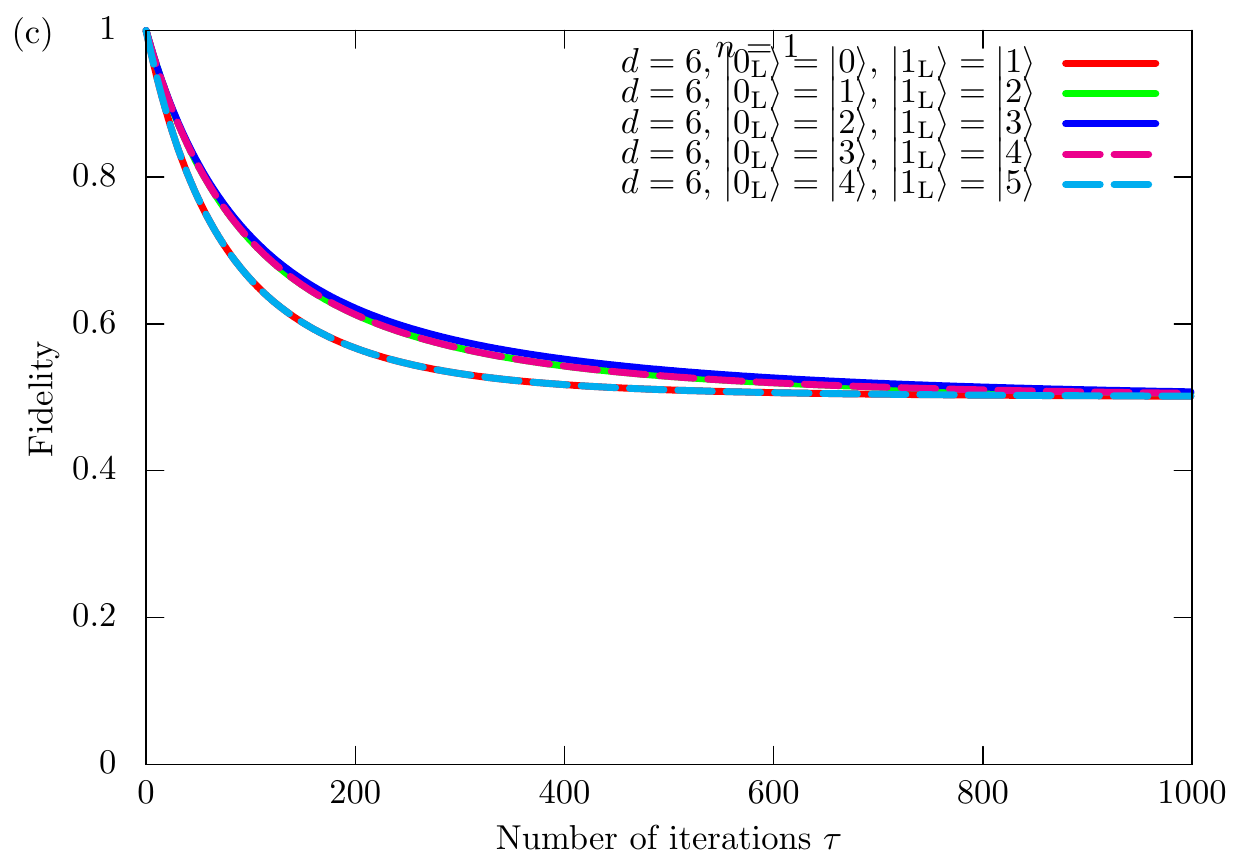}
\includegraphics[scale=0.400]{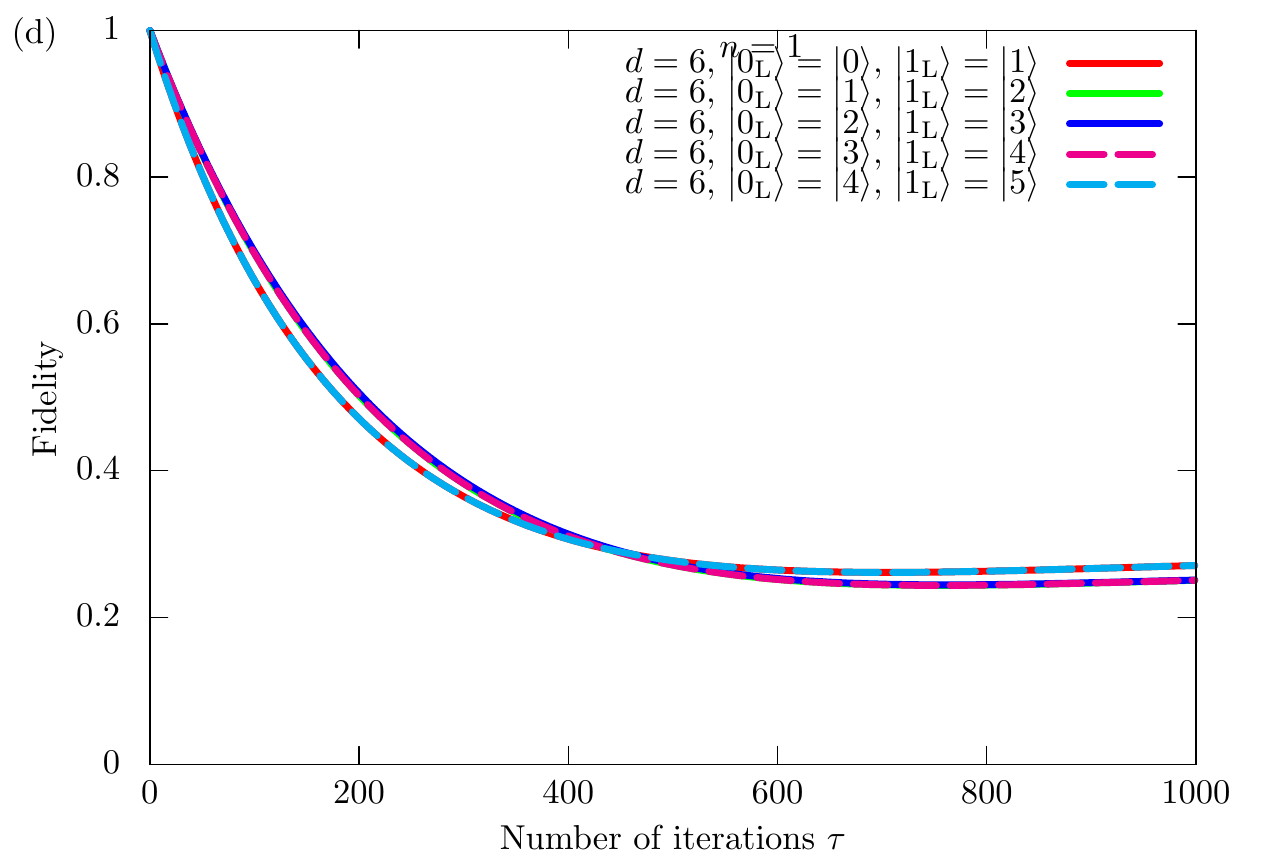}
\caption{Process fidelity for (a) the $Z$-type error model, (b) the $X'$-type error model, and (c) the $X'+Z$-type error model. We have varied $| 0_\mathrm{L} \rangle$ and $| 1_\mathrm{L} \rangle$ simultaneously keeping the distance. The initial state is the maximally entangled state.}
\label{main_fidelity_000_003}
\end{figure}

To investigate this, we fix the value $d=6$ as well as the state $| 0_\mathrm{L} \rangle = |0\rangle$ but vary the distance between two logical states by increasing $| 1_\mathrm{L} \rangle$.
In Fig.~\ref{main_fidelity_000_002}, we show the $| 1_\mathrm{L} \rangle$-dependence of the process fidelity for the $Z$-, $X'$- and $X'+Z$-type error models.
The observed trends are similar to those of Fig.~\ref{main_fidelity_000_001}.  We therefore conclude that the distance between $| 0_\mathrm{L} \rangle$ and $| 1_\mathrm{L} \rangle$ is the key factor, not the dimensionality of the qudit itself.  (Although dimensionality $d$ needs to be high.)
In Figure~\ref{main_fidelity_000_003} we show process fidelity when we vary $| 0_\mathrm{L} \rangle$ and $| 1_\mathrm{L} \rangle$ simultaneously while keeping the distance fixed (set equal to 1).
All curves are nearly identical, meaning that the lifetime does not depend on the choice of sublevels.  This result supports the view that lifetime depends primarily on distance between levels.

\begin{figure}[t]
\centering
\includegraphics[scale=0.400]{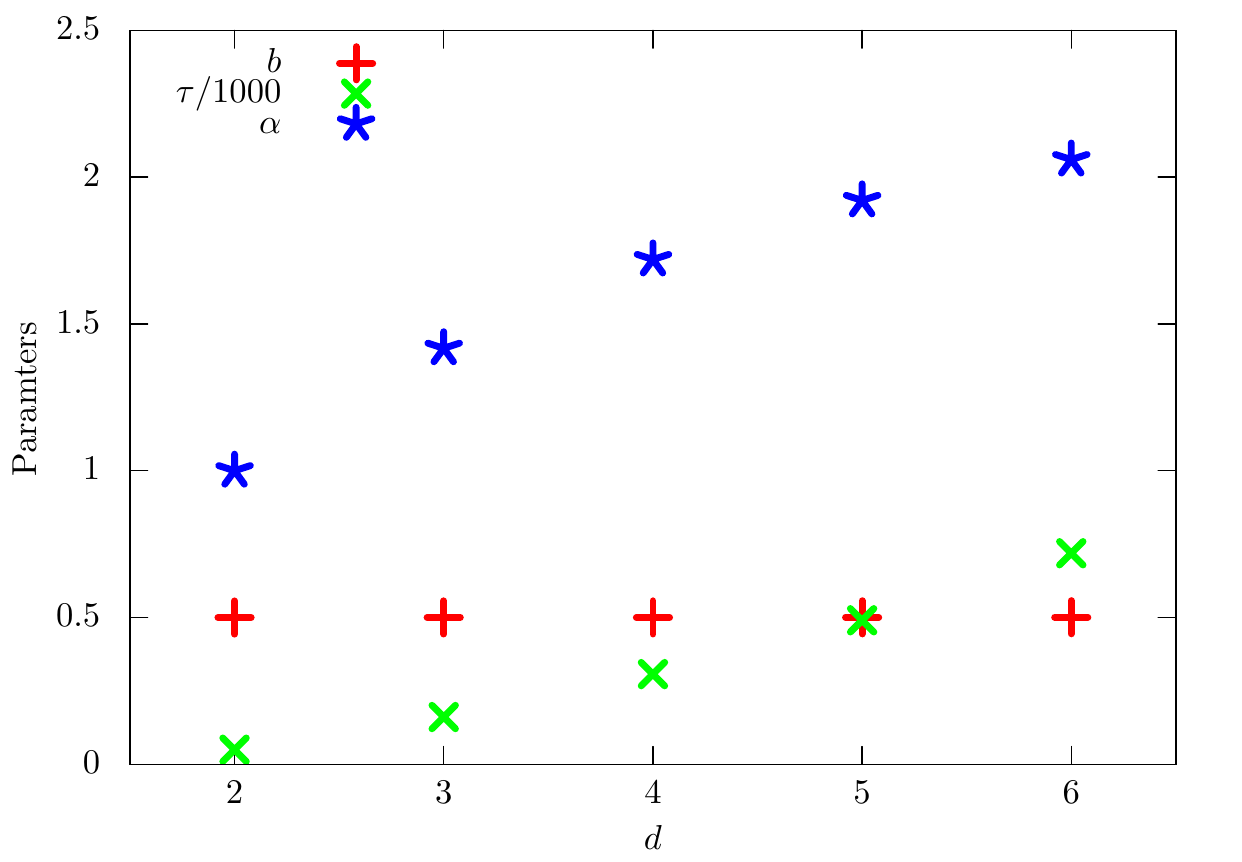}
\includegraphics[scale=0.400]{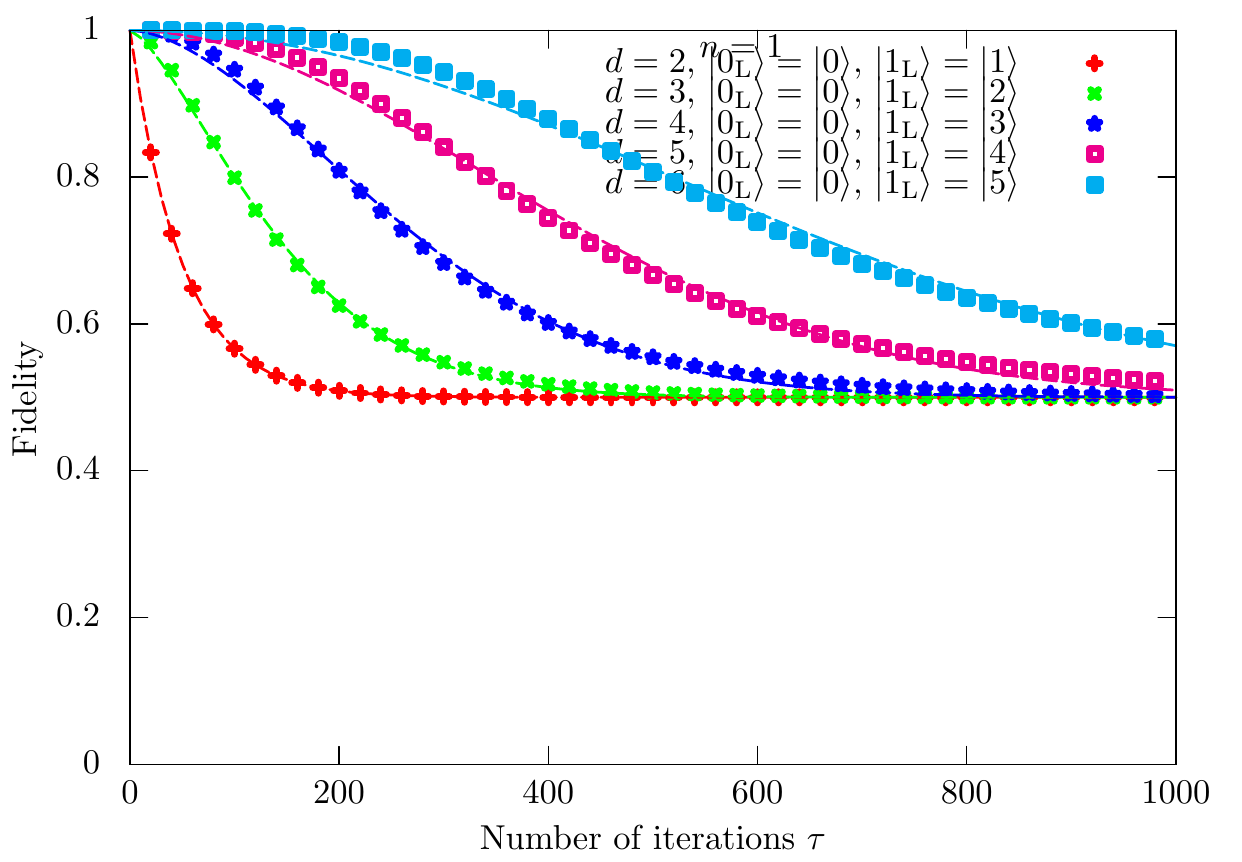}
\caption{(a) Dependence of $b$, $\tau$, and $\alpha$ on $d$. (b) Raw data and fitting curves. The original data are shown in dots and the fitting curves are depicted via dashed lines.}
\label{main_fidelity_000_301}
\end{figure}

Thus, the process fidelity decays slower as the separation between logical levels is increased.  Consider the experiment where $d$ is increased and the logical qubit is encoded in the maximally polarized states. We fit the Kohlrausch (stretched exponential) function ~\cite{Kohlrausch_001, Elton_001}
\begin{align}
f(t;b,\tau,\alpha) &\coloneqq (1-b) e^{-(t/\tau)^\alpha}+b.
\end{align}
for each $d$ and obtain its parameters $\alpha$, $\tau$ and $b$. (A parameter estimation method was proposed in Ref.~\cite{June_001}.)
In Table~\ref{main_table_fidelity_001_001}, the values of $b$, $\tau$, and $\alpha$ obtained from the fit are shown.
\begin{table}[t]
\caption{Estimated values of $b$, $\tau$, and $\alpha$ for each $d$.}
\label{main_table_fidelity_001_001}
\begin{ruledtabular}
\begin{tabular}{cccc}
$d$ & $b$ & $\tau$ & $\alpha$ \\
\hline
2 & 0.500 &  49.498 & 1.000 \\
3 & 0.500 & 161.053 & 1.417 \\
4 & 0.500 & 307.247 & 1.720 \\
5 & 0.500 & 490.644 & 1.921 \\
6 & 0.501 & 718.775 & 2.060 \\
\end{tabular}
\end{ruledtabular}
\end{table}
In Fig.~\ref{main_fidelity_000_301}(a), the results shown in Table~\ref{main_table_fidelity_001_001} is plotted and, in Fig.~\ref{main_fidelity_000_301}(b), the raw values and the fitting curves are shown.
Fig.~\ref{main_fidelity_000_301}(b) shows good agreement between raw data (symbols) and the fitted curve.  The value of $\alpha$ increases from 1 to $\sim 2$, whereas the lifetime $\tau$ increases by a factor of $\sim 15$ (more than an order of magnitude).  This result for $\tau$ is important, as it supports the use of qudits as building blocks for more robust quantum memories.  The result for $\alpha$ indicates that for the $X'$ error model.  This can be understood from the form of the $X'$ operator (Eq.~\ref{main_generalized_Pauli_Xp_001_001}) which, when inserted into the quantum map (Eq.~\ref{main_quantum_map_001_001}), leads to a set of coupled ODEs for the elements of the density matrix that describe a detailed balanced first-order rate process with transitions between neighboring levels.  It is well known that distributions of  rate processes lead to stretched exponential relaxation.

\section{Numerical simulation: entropy production} \label{main_sec_numerical_simulations_entropy_production_001_001}

\begin{figure}[t]
\centering
\includegraphics[scale=0.400]{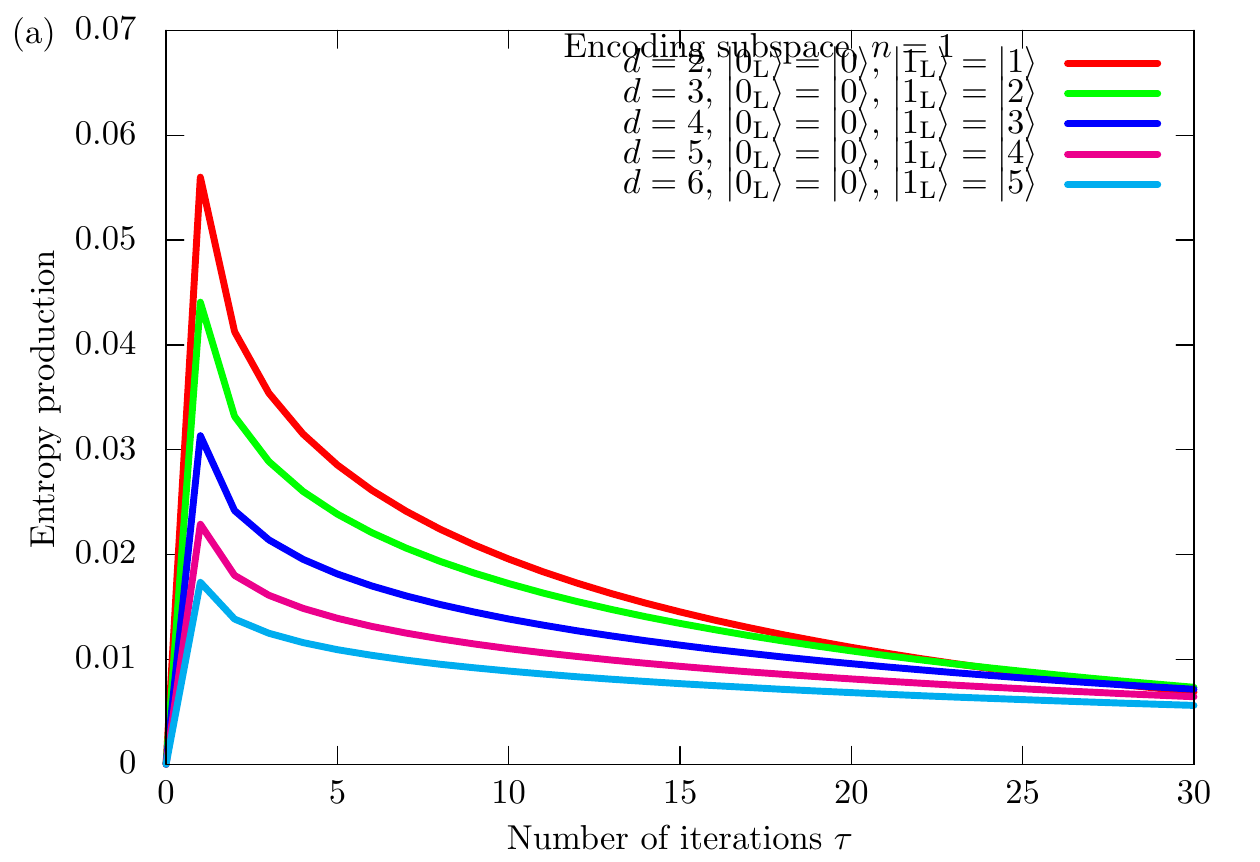}
\includegraphics[scale=0.400]{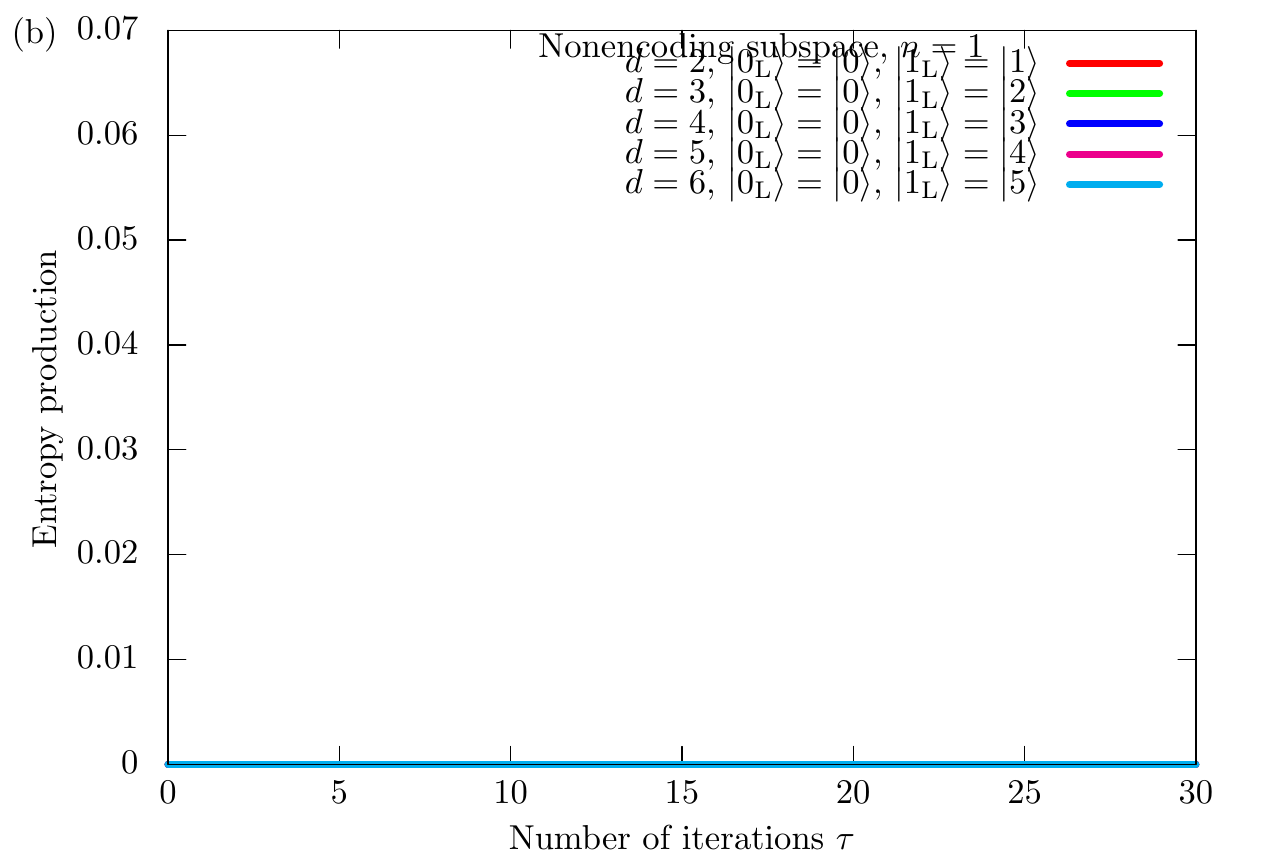}
\includegraphics[scale=0.400]{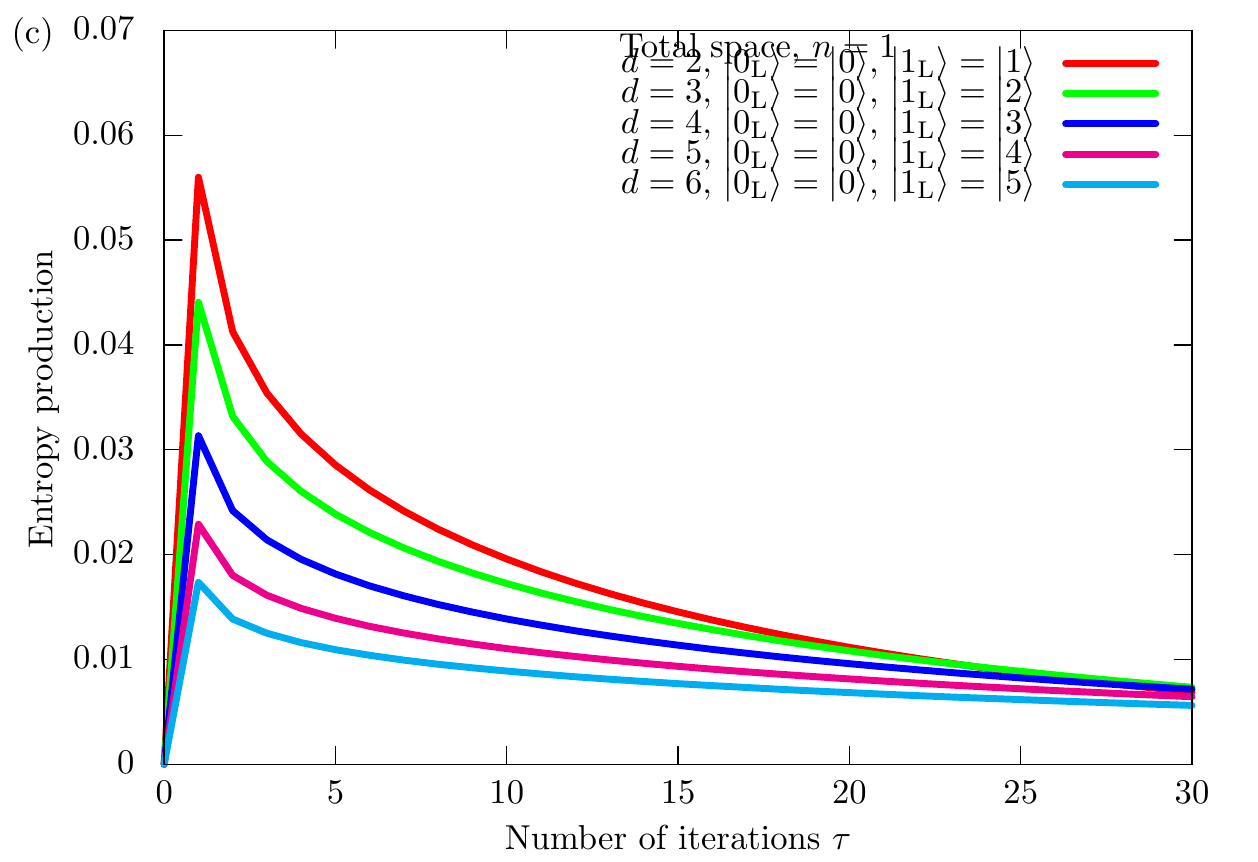}
\caption{Entropy production  of (a) the encoding subspace, (b) the nonencoding subspace, and (c) the total space for the $Z$-type error model. The initial state is the maximally entangled state.}
\label{main_entropy_003_201}
\end{figure}

To shed light on the results of Sec.~\ref{main_sec_numerical_simulations_fidelity_001_001}, we compute entropy production (defined in Sec.~\ref{main_sec_entropy_production_001_001}).  Figure~\ref{main_entropy_003_201} shows entropy production of the total space, the encoding subspace, and the nonencoding subspage for the $Z$-type error model.
We see that in  Fig.~\ref{main_entropy_003_201}(a) entropy production decreases with $d$.
This behavior is consistent with the results of Fig.~\ref{main_fidelity_000_001}(a): shorter lifetimes are associated with high entropy production (rapid loss of information).
The results of  Fig.~\ref{main_entropy_003_201} also show that entropy production in the encoding subspace and the total space are the same but  entropy production in the nonencoding subspace is always zero.
We interpret this to mean that   information initially stored in the logical qubit flows out of the qudit system.  This leakage of information out of the system is less pronounced at large $d$.
% \textcolor{red}{(We should add the explanation from the viewpoint of operators.)}
The lack of entropy production in the nonencoding subspace [Fig.~\ref{main_entropy_003_201}(b)]
is a consequence of the $Z$-type error model, which does not have any channels connecting encoding and nonencoding subspaces (i.e., none of the operators $|k\rangle\langle k|$ present in $Z$ connect pairs of energy levels).

\begin{figure}[t]
\centering
\includegraphics[scale=0.400]{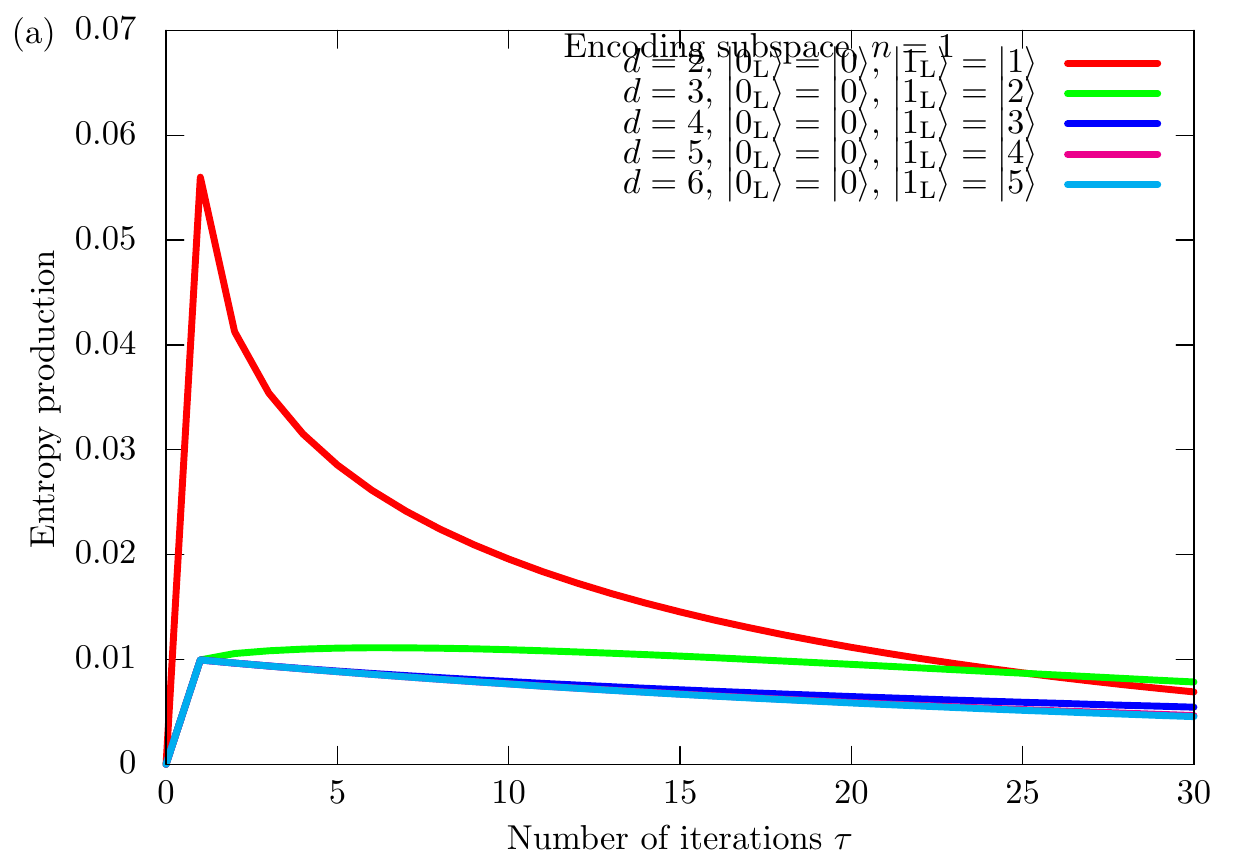}
\includegraphics[scale=0.400]{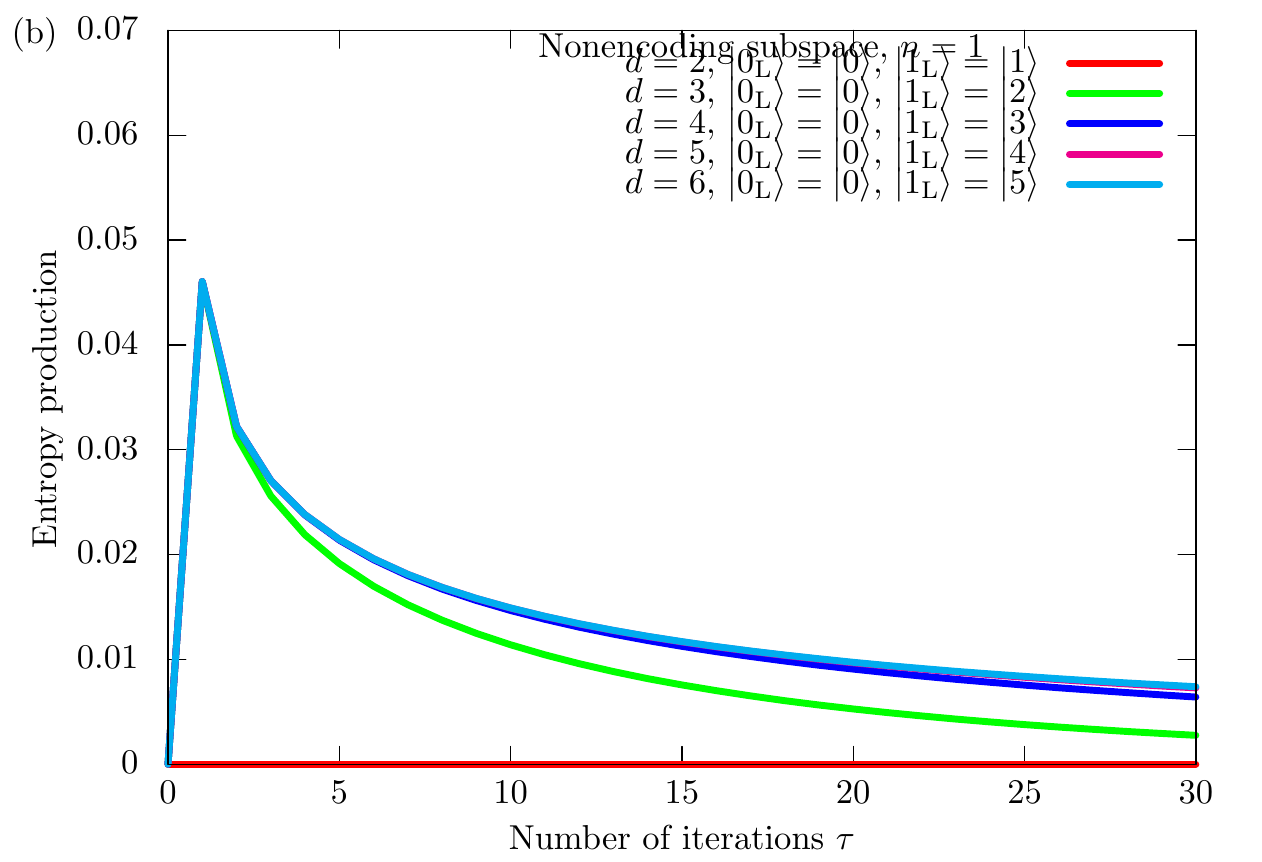}
\includegraphics[scale=0.400]{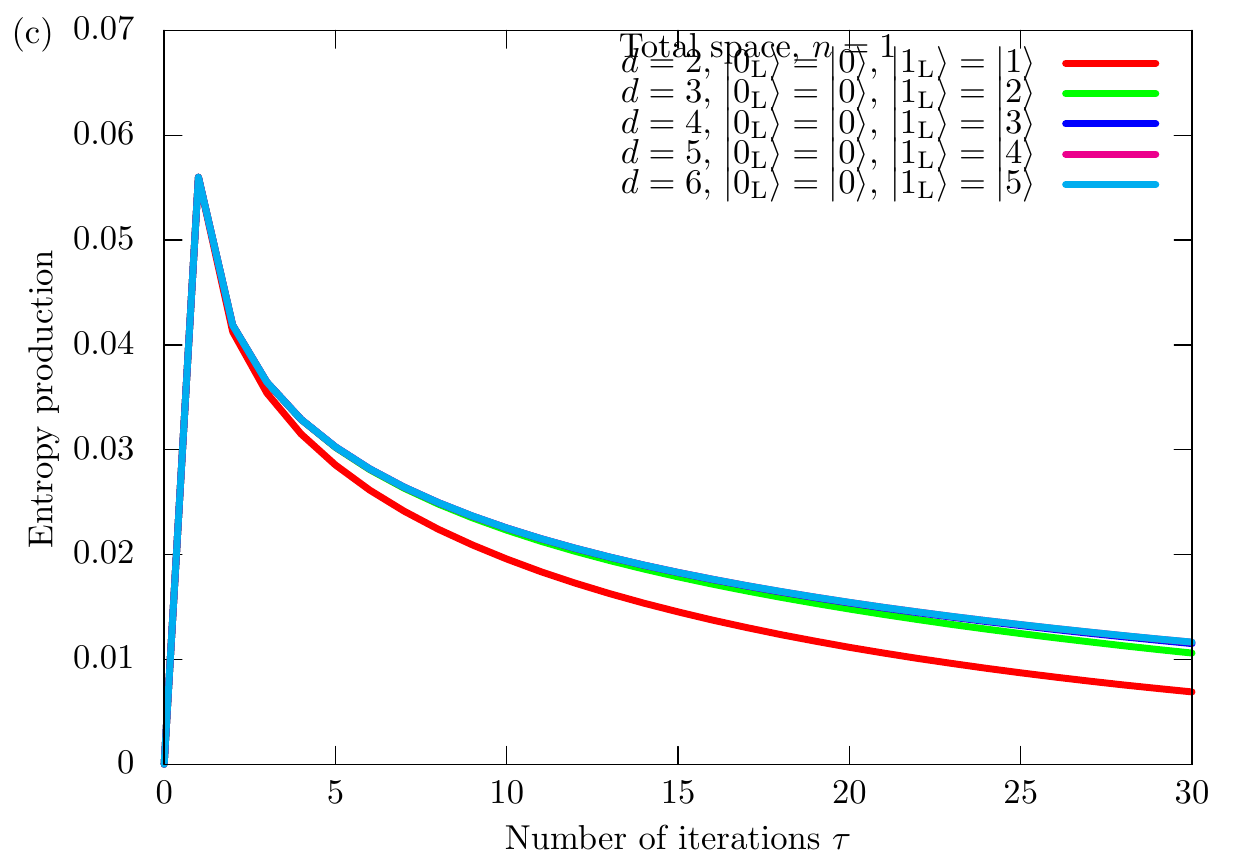}
\caption{Entropy production of (a) the encoding subspace, (b) the nonencoding subspace, and (c) the total space for the $X'$-type error model. The initial state is the maximally entangled state.}
\label{main_entropy_003_202}
\end{figure}

Figure~\ref{main_entropy_003_202} shows entropy production of the total space, the encoding subspace, and the nonencoding for the $X'$-type error model.
In Fig.~\ref{main_entropy_003_202}(a) entropy production within the encoding subspace is drastically suppressed with $d$.
This observation is again consistent with the results of Fig.~\ref{main_fidelity_000_001}(b) (i.e. longer lifetimes with increased $d$ due to the preservation of information).
However, there is an important difference between the cases of the $X'$-type error model and the $Z$-type error model.
For $X'$ the entropy production in the nonencoding subspace is nonzero.  In fact, entropy production in the nonencoding subspace dominates.
This feature enables recovery of qubit information by post-selection.
% \textcolor{red}{(We explain where the entropy is produced. The non-encoding subspace does not exist for $d = 2$.)}
Contrary to the case of the $Z$-type error model, the entropy production in the nonencoding subspace and the total space are nearly identical except for the case $d = 2$.
This leads to increased robustness of the state in the encoding subspace.
We note that for $d = 2$ the nonencoding subpsace does not exist; therefore, $d = 2$ is an important exception.  It is the reason why physical qubits are less ideally suited for use as quantum memories: the only allowed pathway for information transfer is irreversible, whereas for the qudit some of the leakage channels are reversible.

\begin{figure}[t]
\centering
\includegraphics[scale=0.400]{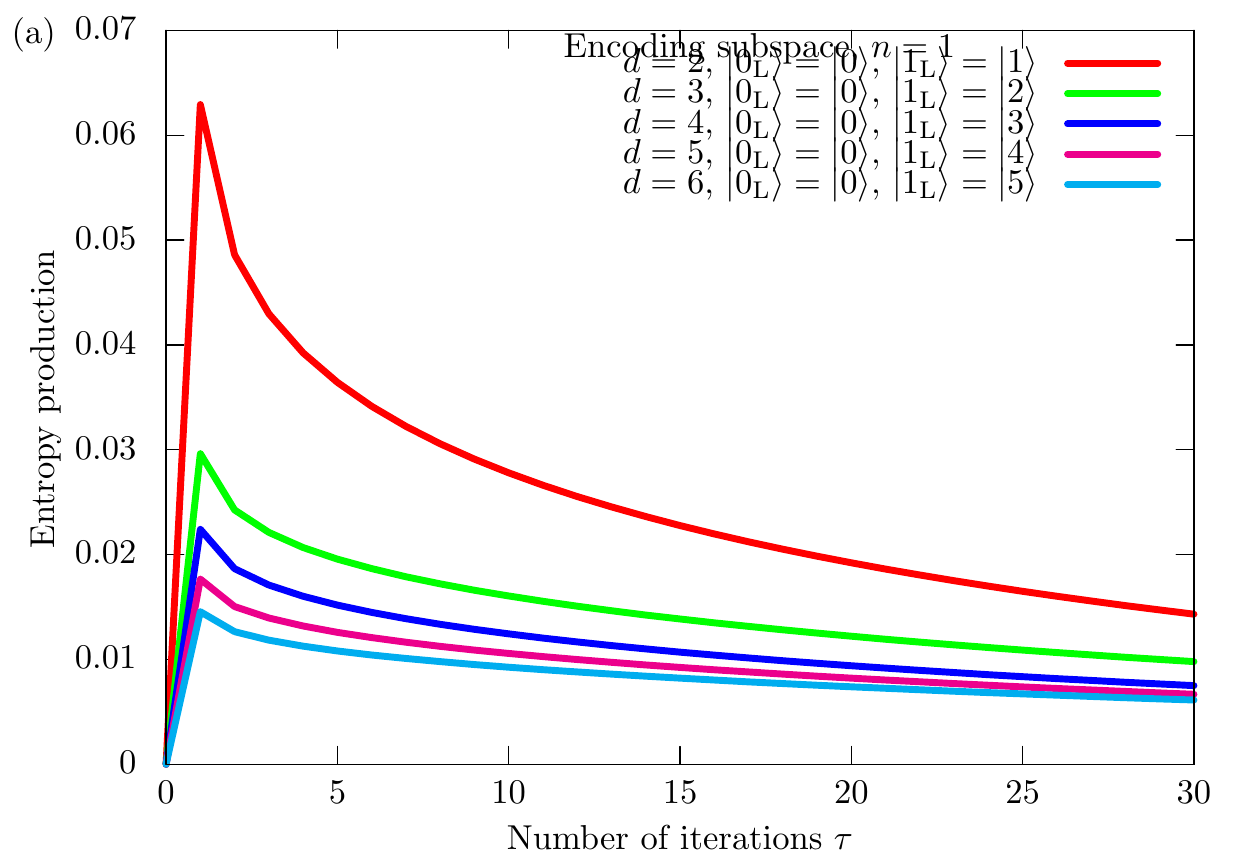}
\includegraphics[scale=0.400]{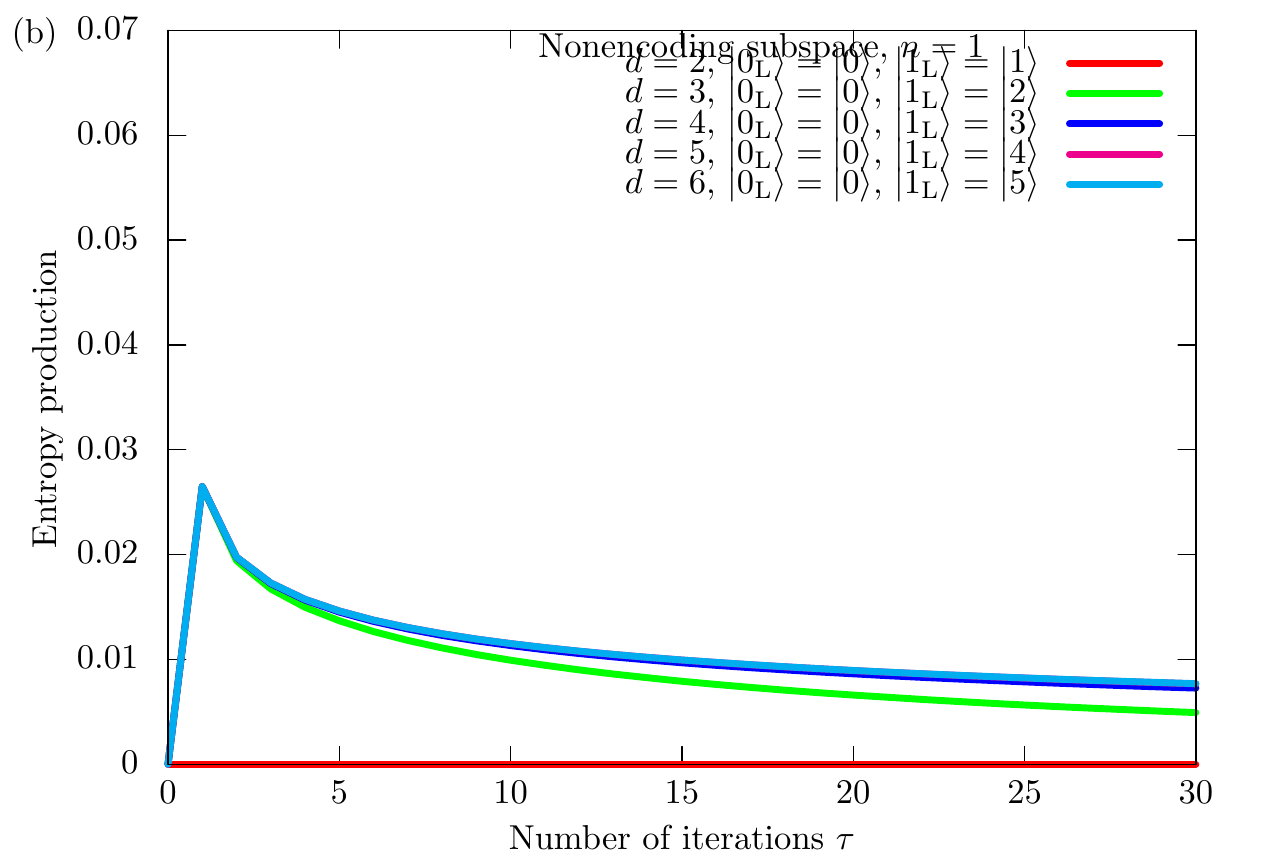}
\includegraphics[scale=0.400]{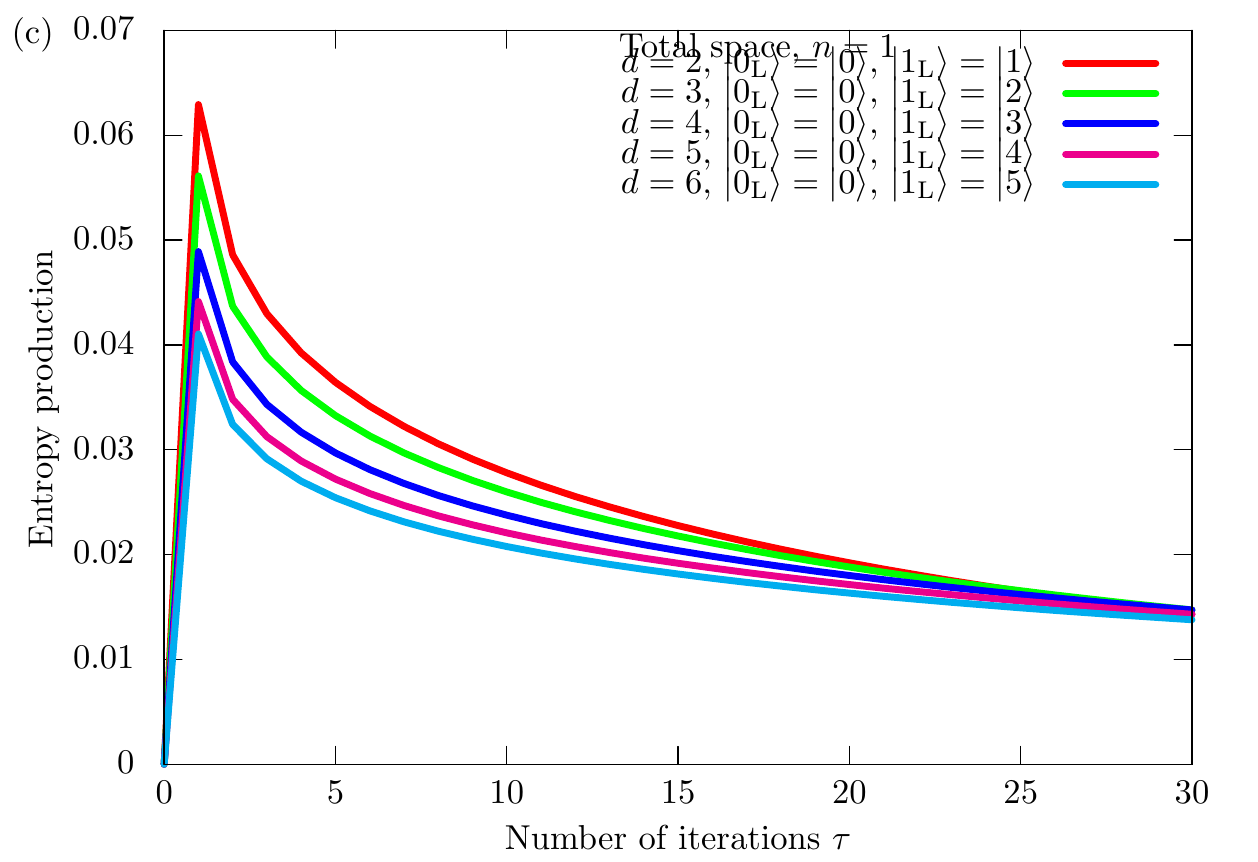}
\caption{Entropy production of (a) the encoding subspace, (b) the nonencoding subspace, and (c) the total space for the $X'$-type error model and $Z$-type error model. The initial state is the maximally entangled state.}
\label{main_entropy_003_203}
\end{figure}

In Fig.~\ref{main_entropy_003_203}, we show entropy production of the total space, the encoding subspace, and the nonencoding for the $X'$-type error model and $Z$-type error model.
In the case of the $X'+Z$-type error model, intermediate behavior  between the above two cases is observed.
Though the improvements become smaller compared to the case of the pure $X'$-type error model, entropy production can still be suppressed by increasing $d$.

\section{Conclusions} \label{main_conclusions_001_001}

We propose the use of qudits as embedding space to encode quantum information from qubits. When the rates of all error channels are kept identical, the resulting quantum memory of the $d$-level system exhibits longer lifetime than the case of physical qubits.  There are, of course, important physical considerations involved when using physical qudits, such as the faster relaxation rates of quadrupolar nuclei or the difficulty of selectively addressing pairs of sublevels.
However, optical qubits~\cite{Wasilewski_001, Mehta_001} and superconducting qubits could be adapted for such applications.
An interpretation of the results was obtained by computing entropy production within the encoding subspace: for the physical qubit, information loss is entirely irreversible whereas for the qudit, information loss from the logical qubit state is only partly irreversible.  The reversible part results in longer lifetimes. We show in Appendix that qudit encoding has properties analogous to  quantum error correction (QEC).  

These results suggest that the encoding of logical qubits in the levels of qudits can provide increased robustness due to the inherent physics of such systems.  Interesting avenues to pursue would be to (i) explore how such physics-aided robustness --- potentially requiring less extra overheads --- compare with the algorithm-aided robustness created in QEC schemes; more specifically, estimating the circuit complexity of a QEC circuit that could afford similar robustness; and (ii) how to integrate the two schemes to obtain simpler circuits and reduced operation overheads that plague direct QEC implementations.  
In order to determine the actual performance gains, a real physical system and its precise characteristics should be modeled. Performance gains over QEC (if any) can only be determined by simulating the precise physical system under consideration, using realistic modeling. A proper QEC scheme for qudits must also be devised; the Appendix presents a simple, but unoptimized one.  Nonetheless, we conclude that these findings are encouraging and warrant further investigations.
This work may motivate and lead to new strategies for quantum processors and quantum memories based on the use of qudits.

\begin{acknowledgments}
We thank Yasunari Suzuki for fruitful discussions.
\end{acknowledgments}

\appendix

\section{Limitations of QEC}

Herein we discuss some limitations of QEC and argue in favor of embedding quantum states inside maximally polarized qudit sublevels as a way to mitigate errors.  The precise benefits of QEC (if any) of course, may depend on the exact nature of the error model used, the precise physical nature of the qudits and available control methods.

\subsection{Fate of maximally entangled states with QEC}

Consider an isolated qudit with embedded logical qubit state (i.e., a quantum memory).   The question we are interested in addressing is whether or not a stored quantum state can survive longer with or without QEC.
Let us consider a simple QEC structure and investigate its properties.

% \subsection{Setup}
%
% \textcolor{red}{(Not finished.)}

\subsubsection{Encoding scheme}

% In Ref.~\cite{Nadkarni_001}, a systematic approach to construct a circuit from stabilizer operators for the repetition code was proposed.
In Ref.~\cite{Nadkarni_001}, a systematic approach to construct a circuit that embeds the measurement values of given stabilizer operators into ancilla qudits for the repetition code was proposed.
% the generalized method of encoding errors that occur in encoding qudits into ancilla qudits are shown.
We review the method.

Suppose that the initial state is $| \psi \rangle$.  Consider the case where we want to embed $n_\mathrm{s}$ stabilizer operators $\{ \hat{S}_i \}_{i=1}^{n_\mathrm{s}}$ into a circuit.
The redundant state, on which the repetition code relies, is given by
\begin{align}
  | \tilde{\psi} \rangle &\coloneqq \prod_{i = 2, 3, \dots, n_\mathrm{r}} \widehat{\text{CNOT}}_{1, i} | \psi \rangle | \underbrace{0 0 \dots 0}_{n_\mathrm{r} - 1} \rangle,
\end{align}
where $n_\mathrm{r}$ is the number of redundant qudits.
% Let us consider the case where we want to embed $n_\mathrm{s}$ stabilizer operators $\{ \hat{S}_i \}_{i=1}^{n_\mathrm{s}}$ into a circuit.
Next, we define the syndrome detection operator $\hat{S}_\mathrm{syn}$:
\begin{align}
  \hat{S}_\mathrm{syn} &\coloneqq \sum_{k_1, k_2, \dots, k_{n_\mathrm{s}} \in \mathbb{F}_d} \hat{S}_1^{k_1} \hat{S}_2^{k_2} \dots \hat{S}_{n_\mathrm{a}}^{k_{n_\mathrm{s}}} \nonumber \\
  & \qquad \qquad \qquad \otimes | k_1 k_2 \dots k_{n_\mathrm{s}} \rangle \langle k_1 k_2 \dots k_{n_\mathrm{s}} |, \label{main_syndrome_operator_001_001}
\end{align}
where $n_\mathrm{r}$ is the number of redundant qudits and $n_\mathrm{a}$ is the number of ancilla qudits.
Note that, as shown in Eq.~\eqref{main_syndrome_operator_001_001}, $n_\mathrm{a} = n_\mathrm{s}$.
% Let us denote an error operator by $\hat{E}$.
By using the following circuit, we can store the measurement values of $\{ \hat{S}_i \}_{i=1}^{n_\mathrm{s}}$ in the ancilla qudits~\cite{Nadkarni_001}:
\begin{align}
\begin{tikzcd}[row sep = 0.2cm, column sep = 0.2cm]
& \lstick{$\hat{E} | \tilde{\psi} \rangle$} & \qw \qwbundle{n_\mathrm{r}}   & \gate[wires=5, nwires=4]{\hat{S}_\mathrm{syn}} & \qw                                   & \qw \rstick{$\hat{E} | \tilde{\psi} \rangle$} \\
& \lstick{$| 0 \rangle$}                    & \gate{\widehat{\mathrm{QFT}}} &                                                & \gate{\widehat{\mathrm{QFT}}^\dagger} & \qw \rstick{$| k_1 \rangle$}                  \\
& \lstick{$| 0 \rangle$}                    & \gate{\widehat{\mathrm{QFT}}} &                                                & \gate{\widehat{\mathrm{QFT}}^\dagger} & \qw \rstick{$| k_2 \rangle$}                  \\
&                                           & \vdots                        &                                                & \vdots                                &                                               \\
& \lstick{$| 0 \rangle$}                    & \gate{\widehat{\mathrm{QFT}}} &                                                & \gate{\widehat{\mathrm{QFT}}^\dagger} & \qw \rstick{$| k_{n_\mathrm{s}} \rangle$}
\end{tikzcd}, \label{main_syndrome_circuit_001_001}
\end{align}
where $\hat{E}$ is an error operator.

% \textcolor{red}{(Not finished.)}

\subsection{3-qudit repetition code and stabilizer group}

% In the previous subsection, we reviewed the general method of constructing a circuit from stabilizer operators.

Consider a 3-qudit repetition code and partial QEC for bit-flipping errors. The initial state of the 3-qudit repetition code is given by
\begin{align}
  | \tilde{\psi} \rangle &\coloneqq \widehat{\text{CNOT}}_{1, 2} \widehat{\text{CNOT}}_{1, 3} | \psi \rangle | 0 \rangle | 0 \rangle,
\end{align}
We consider the following stabilizer group for correct bit-flipping errors:
\begin{align}
  \mathcal{S} &\coloneqq \{ \hat{Z}_1 \hat{Z}_2^\dagger, \hat{Z}_2 \hat{Z}_3^\dagger \}. \label{main_example_qudit_stabilizer_001_001}
\end{align}
Eq.~\eqref{main_example_qudit_stabilizer_001_001} stabilizes the following states:
\begin{align}
  \{ | k, k, k \rangle \}_{k \in \mathbb{F}_d}.
\end{align}
Note that Eq.~\eqref{main_example_qudit_stabilizer_001_001} can correct the bit-flipping errors though it cannot correct all one-qudit errors.

% \textcolor{red}{(Not finished.)}

\subsubsection{Full circuit}

% The whole process is as follows:
In the main text, we clarified that qudits are robust against bit-flipping errors.
Our goal is to correct phase-flipping errors via QEC; Eq.~\eqref{main_syndrome_circuit_001_001} with Eq.~\eqref{main_example_qudit_stabilizer_001_001} can correct them by inserting quantum Fourier transform (QFT) operators before and after the error models.
The full circuit based on Eq.~\eqref{main_syndrome_circuit_001_001} with Eq.~\eqref{main_example_qudit_stabilizer_001_001} is given by
\begin{widetext}
\begin{align}
\begin{tikzcd}[row sep = 0.2cm, column sep = 0.2cm]
& \lstick{$| \psi_\mathrm{ini} \rangle$} & \qw & \ctrl{1} \gategroup[wires=3, steps=2, style={dashed, rounded corners, inner sep= 0pt}]{encoding} & \ctrl{2}  & \qw & \gate{\widehat{\mathrm{QFT}}} & \gate[wires=3]{\text{Error}} & \gate{\widehat{\mathrm{QFT}}^\dagger} & \qw & \qw \gategroup[wires=5, steps=6, style={dashed, rounded corners, inner sep= 0pt}]{correcting} & \gate[wires=5]{\hat{S}_\mathrm{syn}} & \qw                                   & \qw           & \gate{(\hat{X}^\dagger)^{x_1}} & \qw & \qw & \octrl{2} \gategroup[wires=3, steps=2, style={dashed, rounded corners, inner sep= 0pt}]{disentangling} & \octrl{1} & \qw & \qw \rstick{$| \psi_\mathrm{fin} \rangle$} \\
& \lstick{$| 0 \rangle$}                 & \qw & \targ \qw                                                                                        & \qw       & \qw & \gate{\widehat{\mathrm{QFT}}} &                              & \gate{\widehat{\mathrm{QFT}}^\dagger} & \qw & \qw                                                                                           &                                      & \qw                                   & \qw           & \gate{(\hat{X}^\dagger)^{x_2}} & \qw & \qw & \qw                                                                                                    & \targ \qw & \qw & \qw                                        \\
& \lstick{$| 0 \rangle$}                 & \qw & \qw                                                                                              & \targ \qw & \qw & \gate{\widehat{\mathrm{QFT}}} &                              & \gate{\widehat{\mathrm{QFT}}^\dagger} & \qw & \qw                                                                                           &                                      & \qw                                   & \qw           & \gate{(\hat{X}^\dagger)^{x_3}} & \qw & \qw & \targ \qw                                                                                              & \qw       & \qw & \qw                                        \\
& \lstick{$| 0 \rangle$}                 & \qw & \qw                                                                                              & \qw       & \qw & \qw                           & \qw                          & \qw                                   & \qw & \gate{\widehat{\mathrm{QFT}}}                                                                 & \qw                                  & \gate{\widehat{\mathrm{QFT}}^\dagger} & \meter{$m_1$} &                                &     &     &                                                                                                        &           &     &                                            \\
& \lstick{$| 0 \rangle$}                 & \qw & \qw                                                                                              & \qw       & \qw & \qw                           & \qw                          & \qw                                   & \qw & \gate{\widehat{\mathrm{QFT}}}                                                                 & \qw                                  & \gate{\widehat{\mathrm{QFT}}^\dagger} & \meter{$m_2$} &                                &     &     &                                                                                                        &           &     &
\end{tikzcd}, \label{main_whole_circuit_001_001}
\end{align}
\end{widetext}
where
\begin{align}
\begin{tikzcd}[row sep = 0.2cm, column sep = 0.2cm]
& \ctrl{1} & \qw \\
& \targ{}  & \qw
\end{tikzcd} &\coloneqq \widehat{\text{CNOT}}_{1, 2}, \\
\begin{tikzcd}[row sep = 0.2cm, column sep = 0.2cm]
& \octrl{1} & \qw \\
& \targ{}   & \qw
\end{tikzcd} &\coloneqq \widehat{\text{CNOT}}_{1, 2}^\dagger,
\end{align}
and the measurements are performed in the computational basis by using $\hat{P}^k$ for $k \in \mathbb{F}_d$.
Note that $\widehat{\text{CNOT}}_{i, j} \ne \widehat{\text{CNOT}}_{i, j}^\dagger$ unlike the conventional CNOT gate.
% Here, we have used Quantikz~\cite{Kay_001}.

% When $m_2 = 0$, we have $x_1 = m_1$, $x_2 = 0$, and $x_3 = 0$.
% When $m_1 = 0$, we have $x_1 = 0$, $x_2 = 0$, and $x_3 = - m_2 \ (\mathrm{mod} \ d)$.
% When $m_1 = -i \ (\mathrm{mod} \ d)$ and $m_2 = i$, we have $x_1 = 0$, $x_2 = i$, and $x_3 = 0$.
% Otherwise, we cannot correct errors.

The correct operators in Eq.~\eqref{main_whole_circuit_001_001} are specified by $(x_1, x_2, x_3)$ and they are functions of the values of measurements $(m_1, m_2)$.
In Table~\ref{main_table_error_001_001}, we summarize the relationship between $(m_1, m_2)$ and $(x_1, x_2, x_3)$.
\begin{table}[t]
\caption{Relationship between measurements and correction operators. Only correctable patterns of measurements are shown. In the table, $i$ takes $1, 2, \dots, d-1$.}
\label{main_table_error_001_001}
\begin{ruledtabular}
\begin{tabular}{cc|ccc}
$m_1$ & $m_2$ & $x_1$ & $x_2$ & $x_3$ \\
\hline
$i$                       & $0$                       & $i$ & $0$ & $0$ \\
$-i \ (\mathrm{mod} \ d)$ & $i$                       & $0$ & $i$ & $0$ \\
$0$                       & $-i \ (\mathrm{mod} \ d)$ & $0$ & $0$ & $i$ \\
\end{tabular}
\end{ruledtabular}
\end{table}

% \textcolor{red}{(Not finished.)}

\subsubsection{Numerical results}

\begin{figure}[t]
\centering
\includegraphics[scale=0.400]{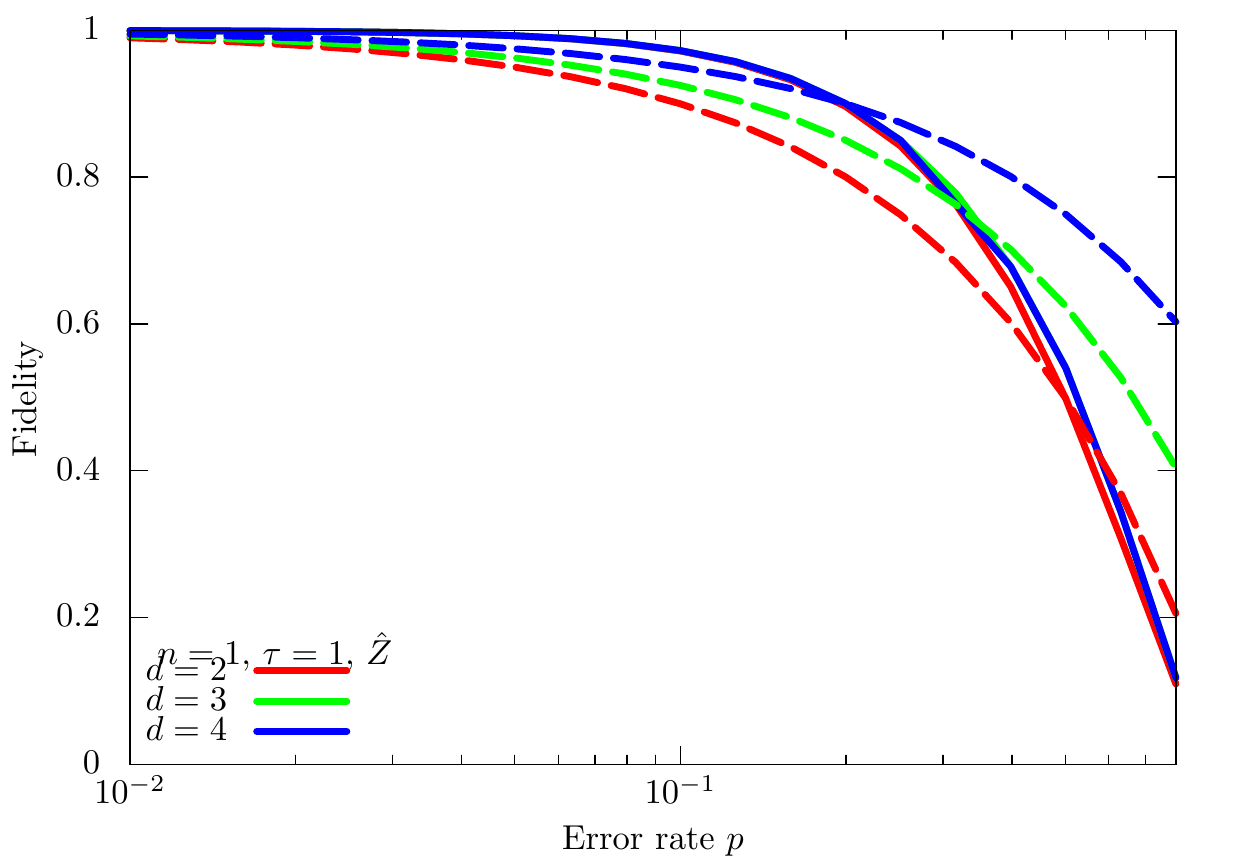}
\includegraphics[scale=0.400]{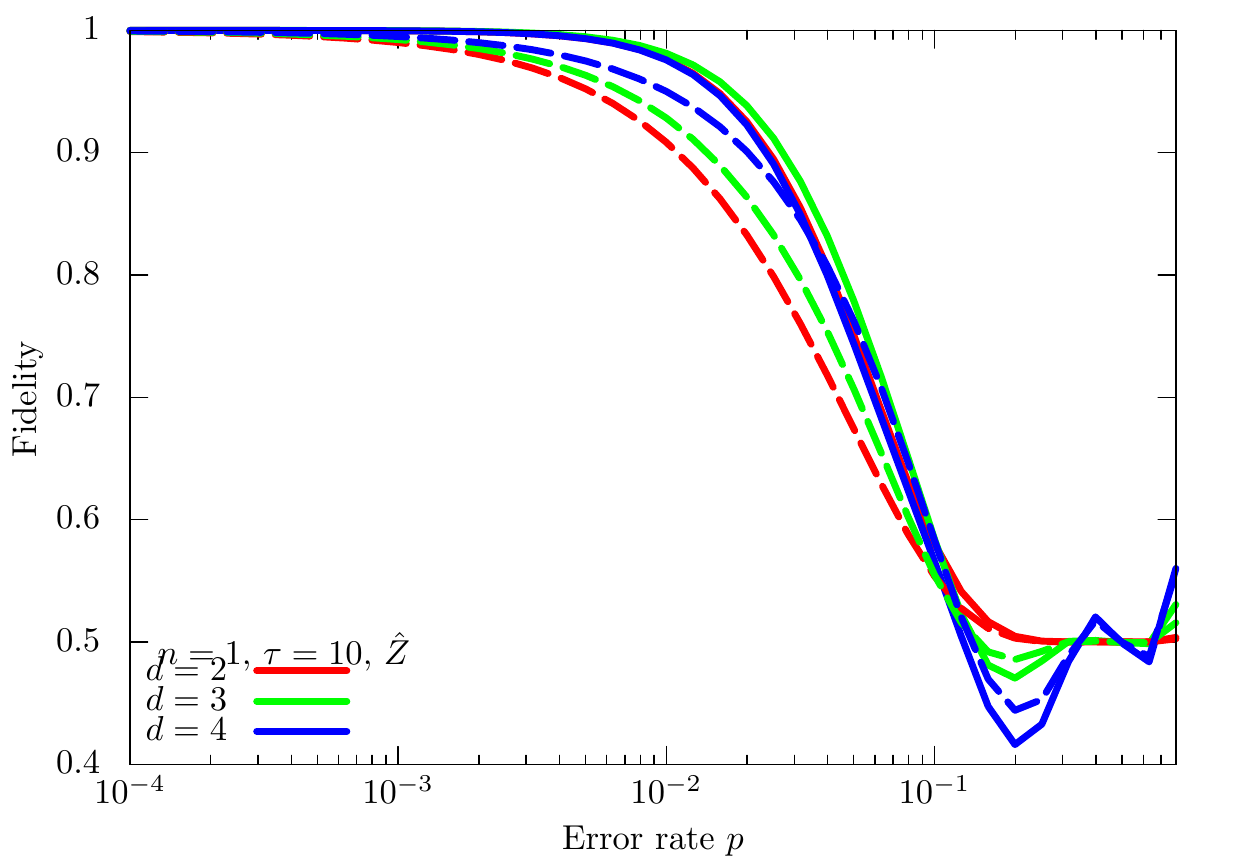}
\includegraphics[scale=0.400]{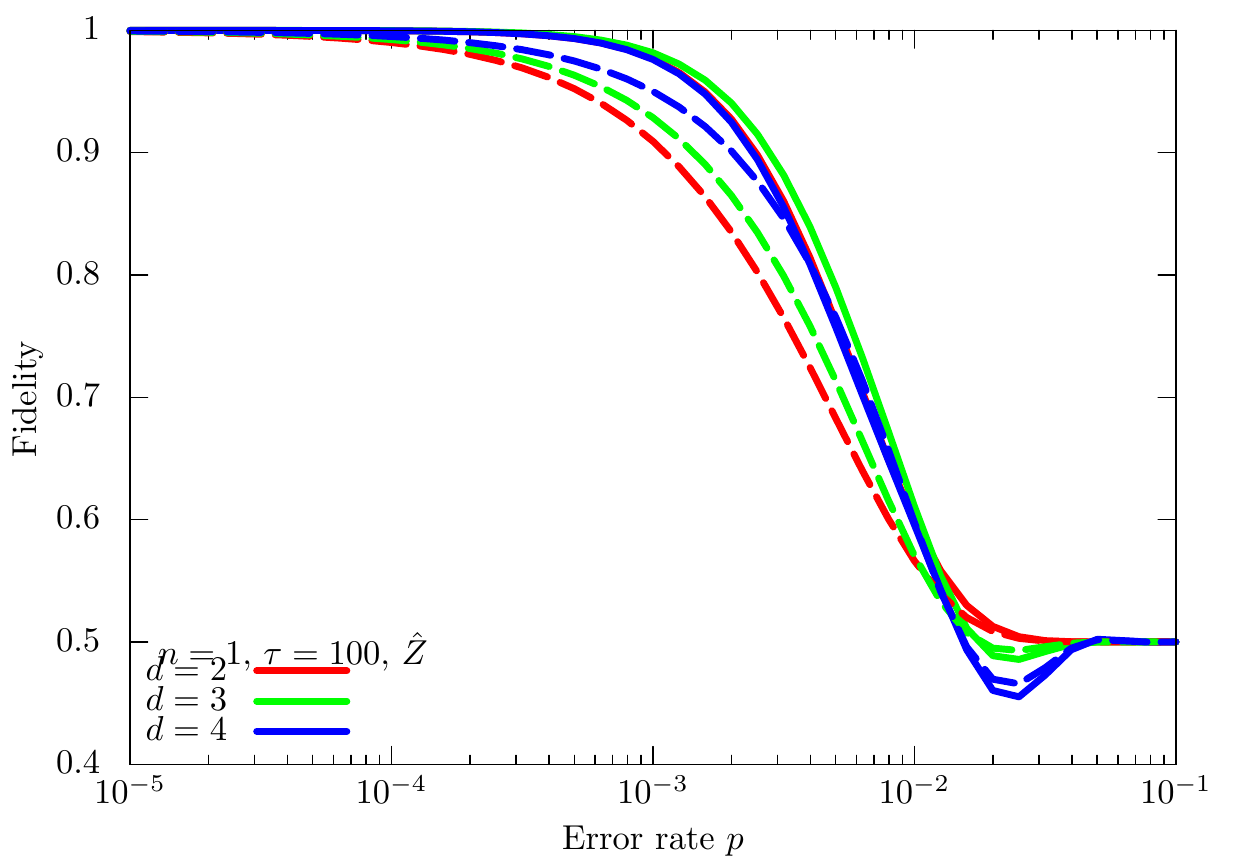}
\caption{Process fidelity for the $Z$-type error model. Solid lines and dashed lines depict results with and without QEC, respectively. We use states at (a) $\tau = 1$, (b) $\tau = 10$, and (c) $\tau = 100$. The initial state is the maximally entangled state.}
\label{main_fidelity_002_001}
\end{figure}

Figure~\ref{main_fidelity_002_001} shows the process fidelity with and without QEC in the case of the $Z$-type error model by solid and dashed lines, respectively.
Since the QEC circuit in Eq.~\eqref{main_whole_circuit_001_001} is designed for phase-flipping errors, the process fidelity with QEC is higher than that without QEC for small $p$.

% In Fig.~\ref{main_trace_002_011},
% \begin{figure}[t]
% \centering
% \includegraphics[scale=0.400]{pic_standalone_002/pic_gnuplot_002_011/main_001_001.pdf}
% \includegraphics[scale=0.400]{pic_standalone_002/pic_gnuplot_002_011/main_001_002.pdf}
% \includegraphics[scale=0.400]{pic_standalone_002/pic_gnuplot_002_011/main_001_003.pdf}
% \caption{Trace of the subspace density operator for the $Z$-type error model. Solid lines and dashed lines depict results with and without QEC, respectively. We use states at (a) $\tau = 1$, (b) $\tau = 10$, and (c) $\tau = 100$. The initial state is the maximally entangled state.}
% \label{main_trace_002_011}
% \end{figure}

\begin{figure}[t]
\centering
\includegraphics[scale=0.400]{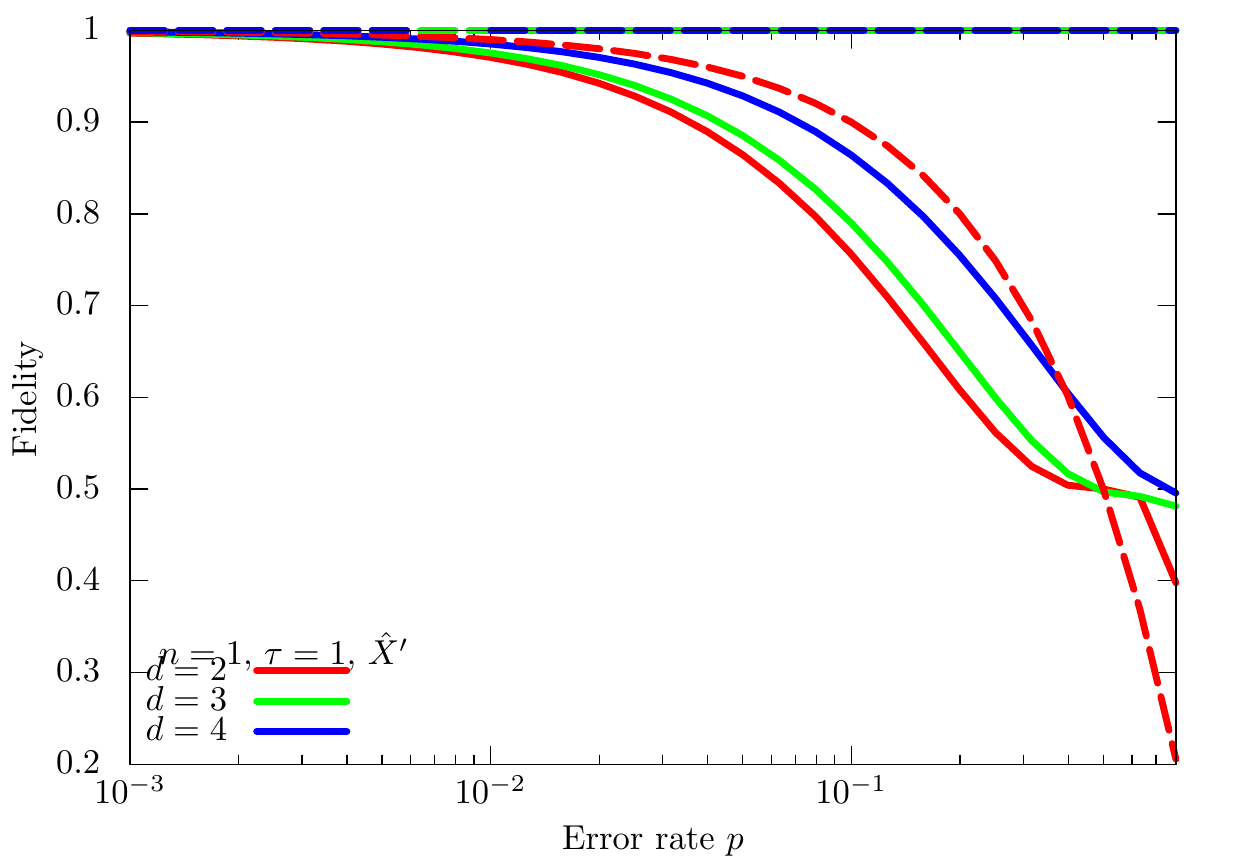}
\includegraphics[scale=0.400]{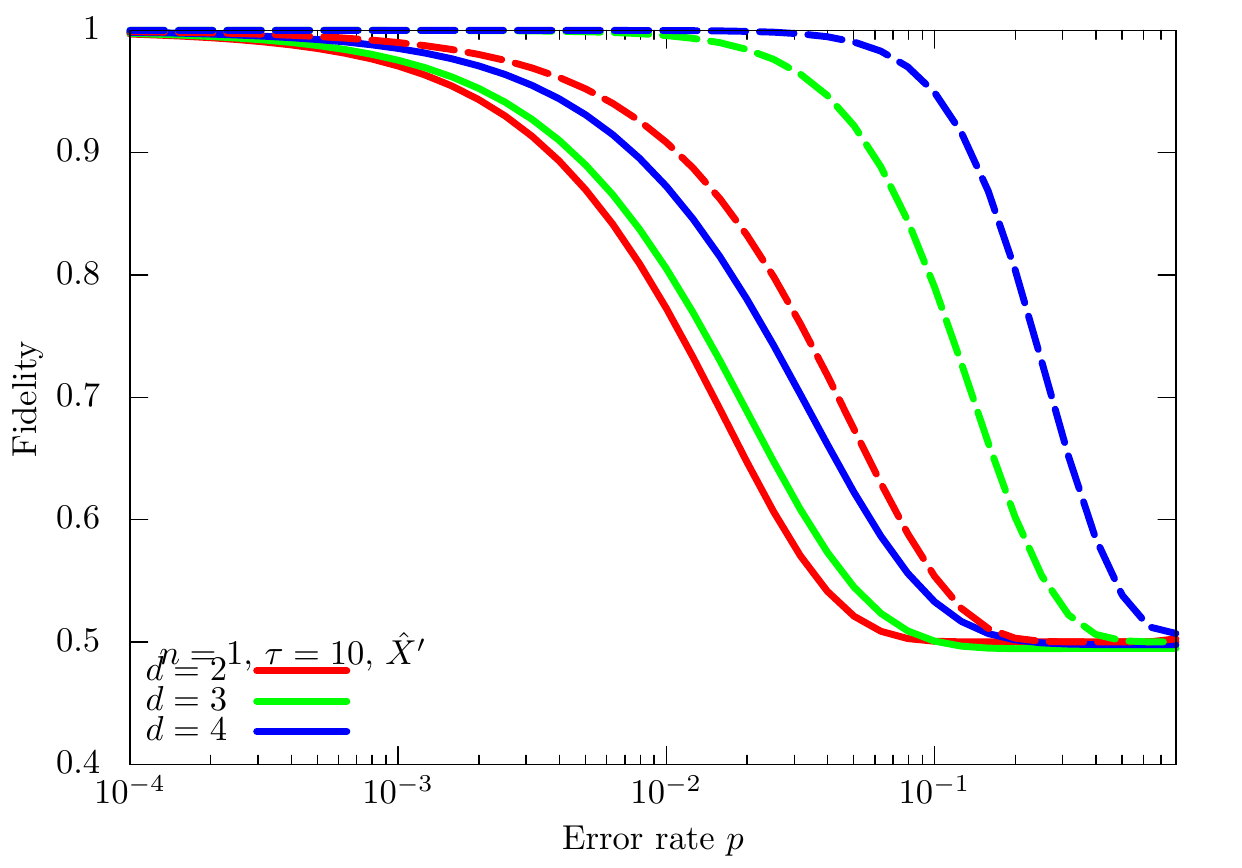}
\includegraphics[scale=0.400]{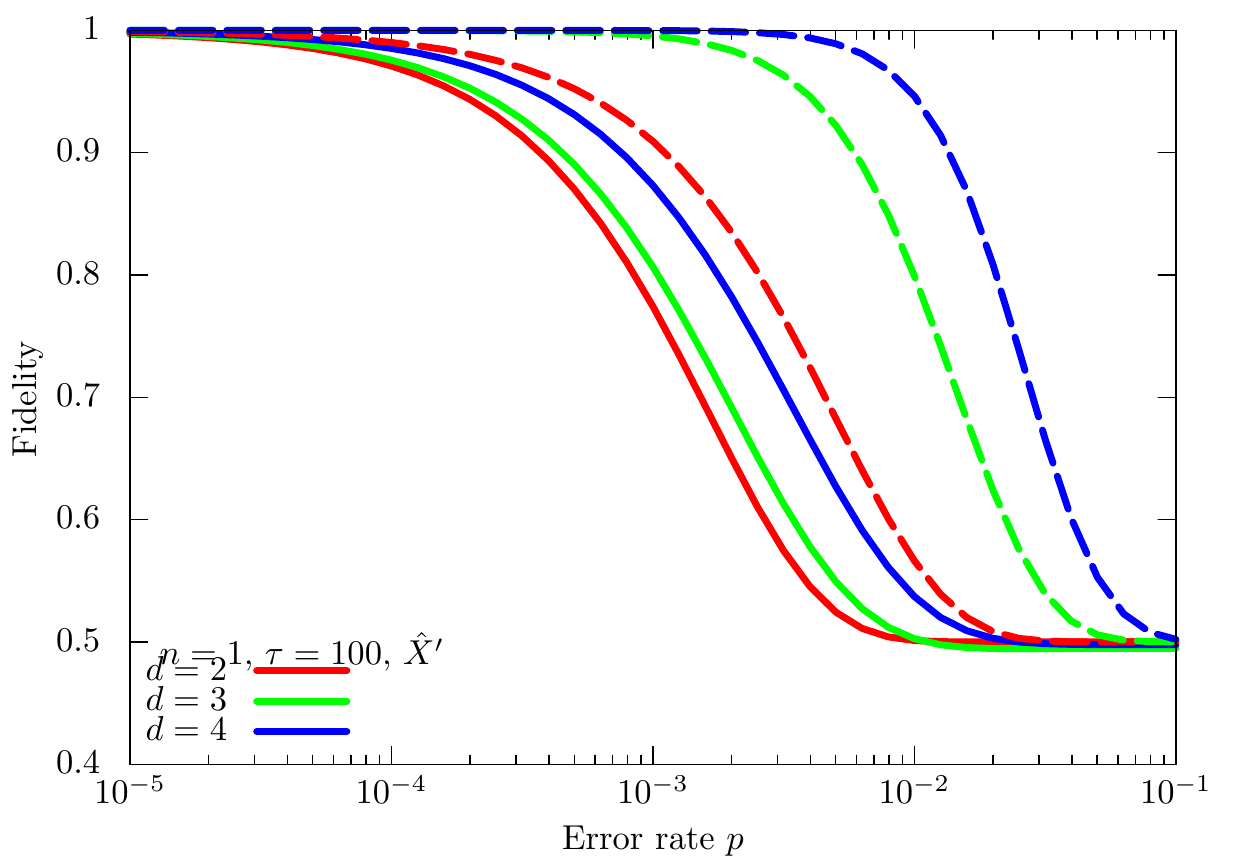}
\caption{Process fidelity for the $X'$-type error model. Solid lines and dashed lines depict results with and without QEC, respectively. We use states at (a) $\tau = 1$, (b) $\tau = 10$, and (c) $\tau = 100$. The initial state is the maximally entangled state.}
\label{main_fidelity_002_002}
\end{figure}

Figure~\ref{main_fidelity_002_002} shows the process fidelity with and without QEC in the case of the $X'$-type error model by solid and dashed lines, respectively.
Since the QEC circuit in Eq.~\eqref{main_whole_circuit_001_001} cannot correct bit-flipping errors, the process fidelity without QEC is higher than that with QEC.

% In Fig.~\ref{main_trace_002_012},
% \begin{figure}[t]
% \centering
% \includegraphics[scale=0.400]{pic_standalone_002/pic_gnuplot_002_012/main_001_001.pdf}
% \includegraphics[scale=0.400]{pic_standalone_002/pic_gnuplot_002_012/main_001_002.pdf}
% \includegraphics[scale=0.400]{pic_standalone_002/pic_gnuplot_002_012/main_001_003.pdf}
% \caption{Trace of the subspace density operator for the $X'$-type error model. Solid lines and dashed lines depict results with and without QEC, respectively. We use states at (a) $\tau = 1$, (b) $\tau = 10$, and (c) $\tau = 100$. The initial state is the maximally entangled state.}
% \label{main_trace_002_012}
% \end{figure}

\begin{figure}[t]
\centering
\includegraphics[scale=0.400]{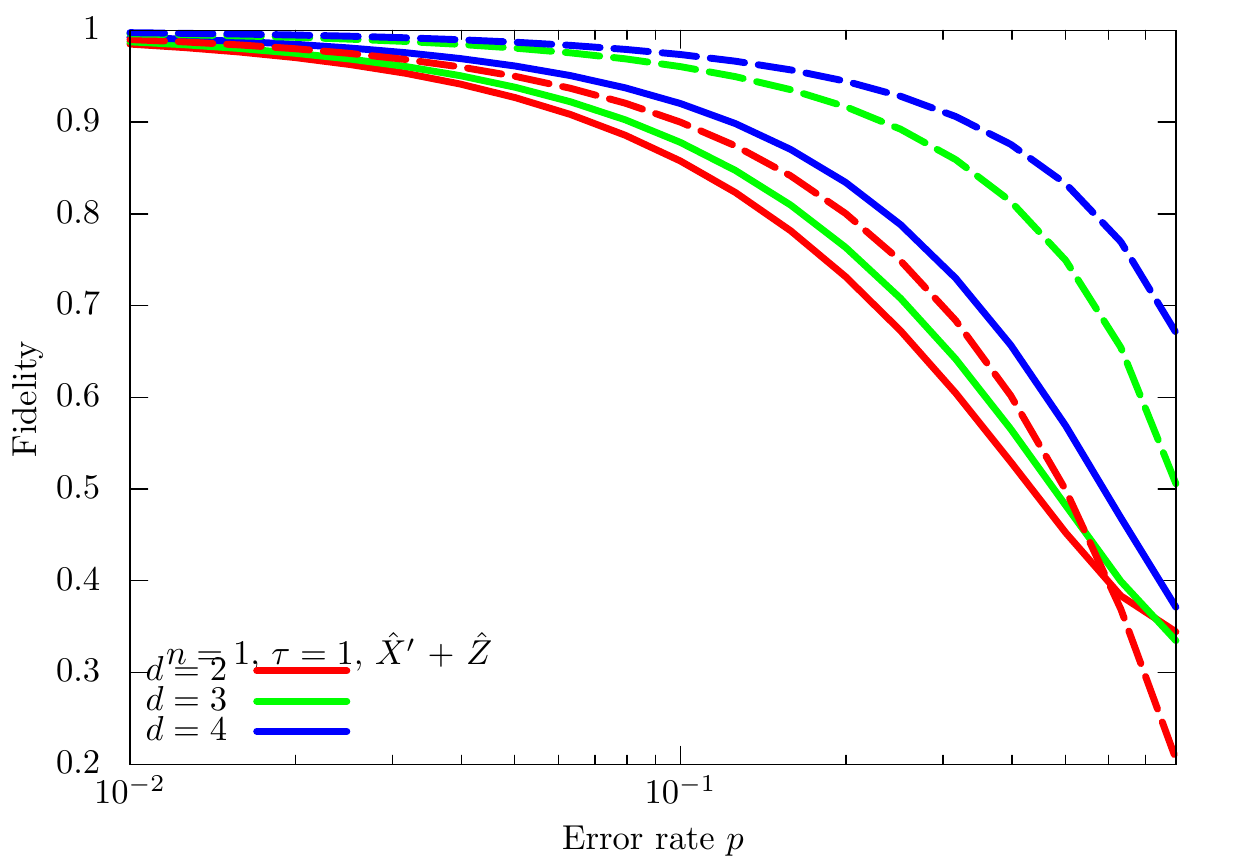}
\includegraphics[scale=0.400]{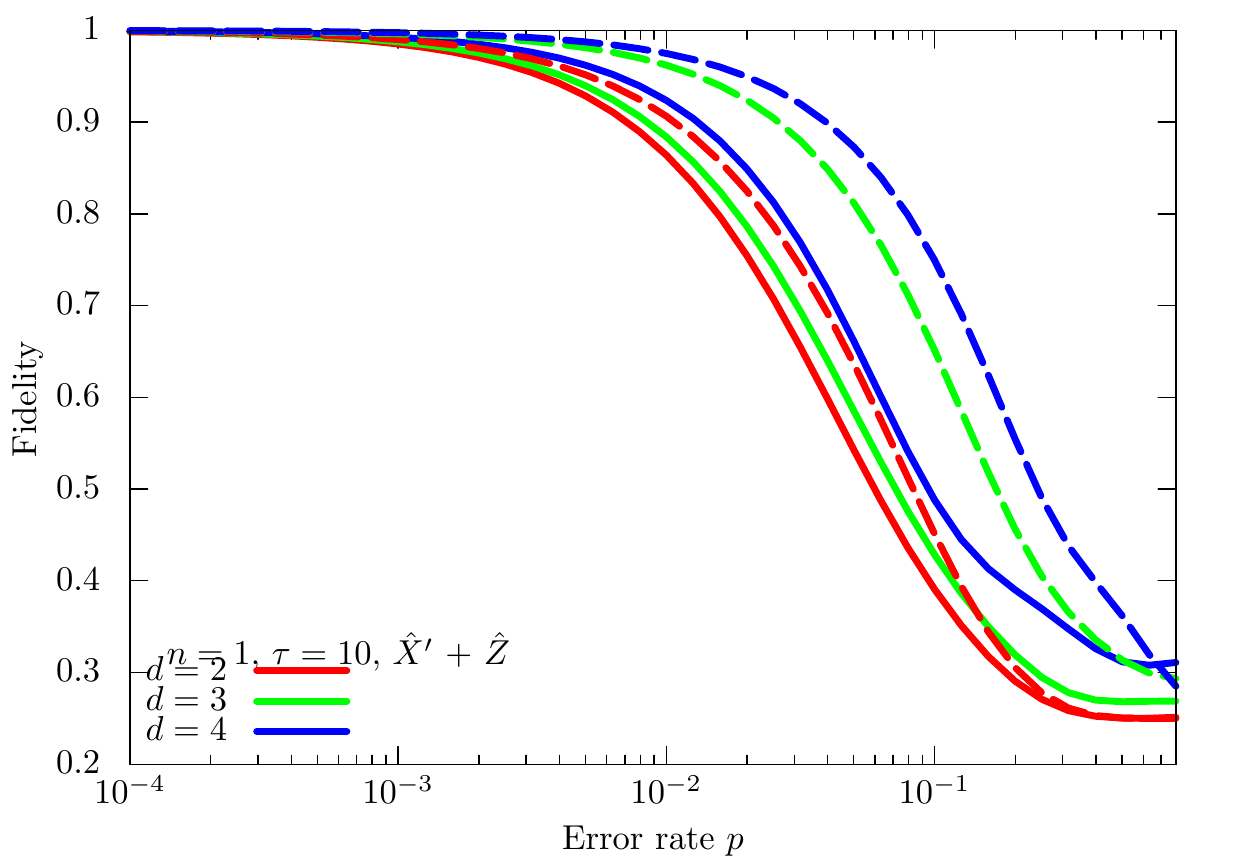}
\includegraphics[scale=0.400]{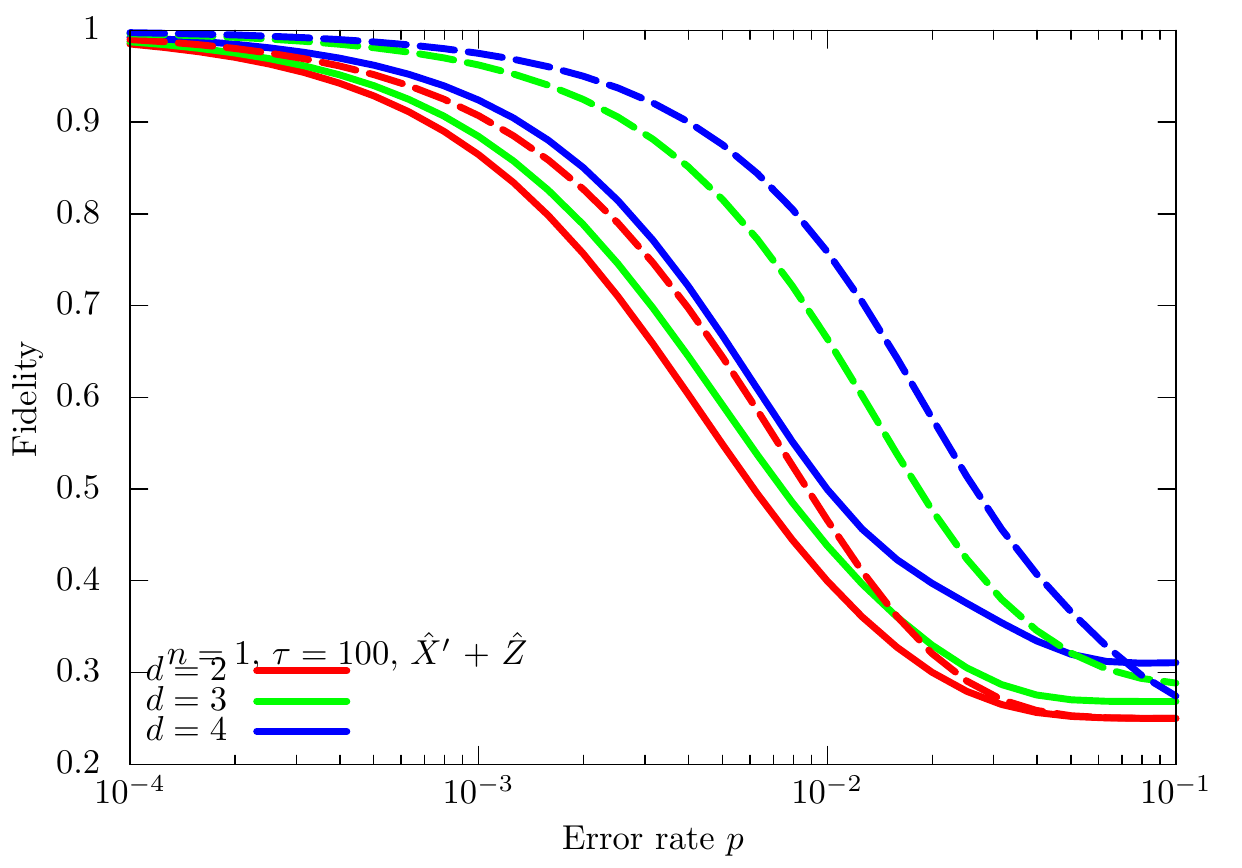}
\caption{Process fidelity for the $X'+Z$-type error model. Solid lines and dashed lines depict results with and without QEC, respectively. We use states at (a) $\tau = 1$, (b) $\tau = 10$, and (c) $\tau = 100$. The initial state is the maximally entangled state.}
\label{main_fidelity_002_003}
\end{figure}

Figure~\ref{main_fidelity_002_003} shows the process fidelity with and without QEC in the case of the $X'+Z$-type error model by solid and dashed lines, respectively.
Figure~\ref{main_fidelity_002_003} shows that, though the QEC circuit in Eq.~\eqref{main_whole_circuit_001_001} can correct phase-flip errors, the process fidelity without QEC is higher than that with QEC.

% In Fig.~\ref{main_trace_002_013},
% \begin{figure}[t]
% \centering
% \includegraphics[scale=0.400]{pic_standalone_002/pic_gnuplot_002_013/main_001_001.pdf}
% \includegraphics[scale=0.400]{pic_standalone_002/pic_gnuplot_002_013/main_001_002.pdf}
% \includegraphics[scale=0.400]{pic_standalone_002/pic_gnuplot_002_013/main_001_003.pdf}
% \caption{Trace of the subspace density operator for the $X'+Z$-type error model. Solid lines and dashed lines depict results with and without QEC, respectively. We use states at (a) $\tau = 1$, (b) $\tau = 10$, and (c) $\tau = 100$. The initial state is the maximally entangled state.}
% \label{main_trace_002_013}
% \end{figure}

% \textcolor{red}{(Not finished.)}

\subsection{Remarks}

Though QEC for both bit-flipping and phase-flipping errors was not investigated in this paper, it is expected that it can improve the process fidelity.
However, to implement QEC for both bit-flipping and phase-flipping errors such as the surface code, a large number of qubits are required.
We should note that the QEC we considered for phase-flipping errors is simple.
Our simulation results show that the process fidelity is improved only for the $Z$-type error model but performance gets worse for $X'$- and $X'+Z$-type error models.
This result suggests that the inherent physical robustness of qudits may be an interesting avenue to pursue in conjunction with or in lieu of error correcting codes, which have higher overhead costs.  In order to determine the actual performance gains, a real physical system and its precise characteristics should be modeled.

% \textcolor{red}{(Not finished.)}

% \bibliographystyle{apsrev4-1}
% \bibliographystyle{unsrt}
\bibliography{manuscript}

\end{document}